\documentclass{psp-book9x6-modified}

\usepackage{psp-book-van-modified} 
\usepackage{cite,graphicx,color,amsmath,booktabs,microtype,afterpage,subfigure,fixmath,truncate,multirow}
\usepackage{mathpazo} 
\usepackage[scaled]{helvet} 
\usepackage{bibunits} 
\usepackage{perpage} 
\usepackage[geometry]{ifsym}
\usepackage[colorlinks,plainpages=false,linkcolor=black,urlcolor=black,citecolor=black,pdfpagemode=UseNone,pdfstartview=FitBH]{hyperref}
\usepackage[dvipsnames]{xcolor}
\usepackage[T1]{fontenc}\usepackage[latin1]{inputenc}

\let\Gamma\varGamma
\let\Delta\varDelta
\let\Theta\varTheta
\let\Lambda\varLambda
\let\Xi\varXi
\let\Pi\varPi
\let\Sigma\varSigma
\let\Upsilon\varUpsilon
\let\Phi\varPhi
\let\Psi\varPsi

\makeatletter
\renewcommand\thefigure{\thechapter.\@arabic\c@figure}
\makeatother

\usepackage{perpage}\MakePerPage{footnote}

\defaultbibliographystyle{inosov-book}

\copyline{Rare-Earth Borides}{Dmytro S. Inosov (ed.)}{2020}
\title{Rare-Earth Borides} 

\graphicspath{{Figures/}} 

\begin{document}




\cleardoublepage
\setcounter{page}{1}
\setcounter{chapter}{2}


\chapter[\bf Crystal structures of dodecaborides: \mbox{complexity\hspace{-1em}} \hspace{1em}in simplicity]{\bf Crystal structures of dodecaborides: complexity in simplicity\label{Chapter:Bolotina}}
\addtocontents{toc}{\vspace{-2pt}\hspace{2.7em}\textit{by Nadezhda~B.~Bolotina, Alexander~P.~Dudka,\\\hspace{2.7em}Olga~N.~Khrykina, and Vladimir~S.~Mironov}\smallskip}

\chapauth{Nadezhda~B.~Bolotina$^{\text{a},\ast}$, Alexander~P.~Dudka$^{\text{a,b}}$, Olga~N.~Khrykina$^{\text{a,b}}$, and Vladimir~S.~Mironov$^{\text{a}}$
\chapaff{\noindent
$^\text{a}$Shubnikov Institute of Crystallography of the Federal Scientific Research Center ``Crystallography and Photonics'' of the Russian Academy of Sciences, Leninsky Prospekt 59, 119333 Moscow, Russia\\
$^\text{b}$Prokhorov General Physics Institute, Russian Academy of Sciences, Vavilova str. 38, 119991 Moscow, Russia\\
$^\ast$E-mail address: \href{mailto:bolotina@ns.crys.ras.ru}{bolotina@ns.crys.ras.ru}}}

\begin{bibunit}

\section*{Abstract}

Analysis of the intriguing physical properties of the dodecaborides, $R$B$_{12}$, requires accurate data on their crystal structure. We show that a simple cubic model fits well with the atomic positions in the unit cell but cannot explain the observed anisotropy in the physical properties. The cooperative Jahn-Teller (JT) effect slightly violates the ideal metric of the cubic lattice and the symmetry of the electron density distribution in the lattice interstices. Theoretical models of the JT distortions of the boron framework are presented. Their correspondence to the electron-density distribution on the maps of Fourier syntheses obtained using x-ray data and explaining the previously observed anisotropy of conductive properties is demonstrated. The effect of boron isotope composition on the character of the lattice distortions is shown. We also discuss the application of the Einstein model for cations and the Debye model for the boron atoms to describe the dynamics of the crystal lattice.

\section{Introduction}

\begin{figure}[b]
\begin{center}
\includegraphics[width=0.8\textwidth]{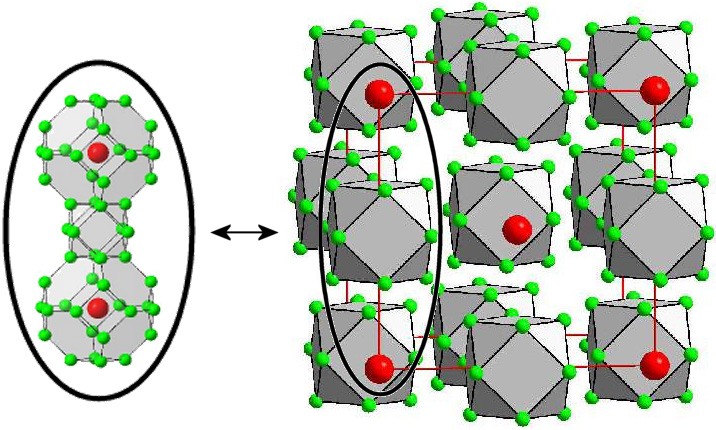}\vspace{-3pt}
\end{center}\index{UB$_{12}$!crystal structure}
\caption{Left:~two metal-centered truncated B$_{24}$ octahedra of the UB$_{12}$-type structure, connected by one empty B$_{12}$ cuboctahedron. Right:~The same structure presented as a NaCl-type structure. The metal atoms (large spheres) alternate in a checkerboard pattern with the B$_{12}$ cuboctahedra.}
\label{Fig:Bolotina_UB12-structure}
\end{figure}

There is a wide variety of borides formed by different-in-shape boron polyhedra in combination with most metals, resulting in a metal-to-boron ratio ranging from 4:1 to 1:66. A highly symmetrical $Fm\overline{3}m$ structure of the uranium dodecaboride UB$_{12}$ was the first one of this type \cite{Bol_BertautBlum49}. A unit cell of the cubic lattice with the lattice constant $a = 7.473$~\AA\ contains four UB$_{12}$ formula units. The metal atoms occupy centers of truncated B$_{24}$ octahedra with boron atoms at each of their 24 vertices. Every boron atom is bonded to two metal atoms and to five other boron atoms. The metal and boron atoms are in positions $4a$~$\{0, 0, 0\}$ and $48i$~$\{0.5, y, y\}$ of the $Fm\overline{3}m$ group, respectively, with $y$ close to 1/6. This compound may also be described in terms of the NaCl-type structure, in which metal atoms and regular B$_{12}$ cuboctahedra occupy the Na and Cl positions, respectively, resulting in a face-centered cubic (fcc) structure shown in Fig.~\ref{Fig:Bolotina_UB12-structure}.

Three years later, the isomorphous ZrB$_{12}$ was prepared and studied \cite{Bol_PostGlaser52}. Six UB$_{12}$-type structures of the rare-earth dodecaborides $R$B$_{12}$ \mbox{($R$~=~Y, Dy, Ho, Er, Tm, Lu)} were determined~\cite{Bol_LaPlacaBinder61} based on the x-ray powder diffraction data. The authors of Ref.~\cite{Bol_PrzybylskaReddoch63} reported the cubic symmetry after examination of a powder ScB$_{12}$ sample. The cubic structure of YB$_{12}$ was later confirmed on single crystals \cite{Bol_MatkovichEconomy65}, but a single crystal of ScB$_{12}$ studied in the same work \cite{Bol_MatkovichEconomy65} was determined as tetragonal \mbox{$I4/mmm$}, with unit-cell parameters \mbox{$a_\text{tet} \approx 5.22$~\AA}, \mbox{$c_\text{tet} \approx 7.35$~\AA} that could be transformed to pseudo-cubic: $a = b = a_\text{tet}\sqrt{2} \approx 7.38$~\AA, $c = c_\text{tet} \approx 7.35$~\AA. All the above findings were summarized in a review article \cite{Bol_MatkovichEconomy77}. A more recent review~\cite{Bol_FlachbartAlekseev08}, which was mainly devoted to magnetic, superconducting, and other physical properties of rare-earth dodecaborides, began by describing the dodecaboride structure and by presenting structural information on $R$B$_{12}$ with $R$~=~Tb\,--\,Lu from the second half of the lanthanide series supplemented with YB$_{12}$ and ZrB$_{12}$. Background information on higher borides of rare earths, including dodecaborides, has been summarized in \cite{Bol_Mori08}. It is worth noting that the rare-earth dodecaborides $R$B$_{12}$ differ from other higher borides $R$B$_n$, $n > 12$. All the dodecaborides, except YbB$_{12}$, are good metals similar to $R$B$_6$ and $R$B$_4$, whereas $R$B$_n$ with $n > 12$ are insulators. Both B$_{12}$ cuboctahedra and icosahedra are electron-deficient by two electrons. The trivalent rare-earth atoms can supply three electrons, so there is one excess conduction electron per unit cell. The only exception is the narrow-gap semiconductor YbB$_{12}$, known also as a Kondo insulator, where Yb takes an intermediate valence.\index{intermediate valence}\index{fluctuating valence}\index{mixed valence} The summary table of structural, electronic and magnetic characteristics of the dodecaborides of the UB$_{12}$ type different in isotope boron composition is presented in Ref.~\cite{Bol_WerheitFilipov11}. In addition to the rare-earth dodecaborides TbB$_{12}$\,--\,LuB$_{12}$ mentioned above, this table contains ZrB$_{12}$, HfB$_{12}$, pseudo-cubic ScB$_{12}$, GdB$_{12}$ synthesized under high pressure \cite{Bol_CannonCannon77} as well as dodecaborides of heavy metals ThB$_{12}$, UB$_{12}$, NpB$_{12}$, PuB$_{12}$ and several solid solutions $R1_xR2_{1-x}$B$_{12}$ of the UB$_{12}$ type. In the same article \cite{Bol_WerheitFilipov11}, the lattice constants of $R$B$_{12}$ as well as the B-B distances in and between the B$_{12}$ cuboctahedra versus ionic radii of the metal atoms are plotted and discussed. A large amount of the reference data on the structures and properties of higher borides, including dodecaborides $R$B$_{12}$, is contained in the overview chapter of a recent PhD thesis \cite{Bol_Akopov18}. Metal dodecaborides $M$B$_{12}$ attract particular interest as multifunctional materials. For instance, in contrast to conventional superhard materials like diamond, which are insulators or semiconductors, many dodecaborides are superhard compounds with high electrical conductivities that can be used as conductors at extreme conditions \cite{Bol_LiangZhang19}.

Owing to the simple cubic structure, dodecaborides are convenient objects for studying physical properties of the metal atoms possessing relative freedom in the large B$_{24}$ cavities of the boron framework. There is a large number of publications on the physical properties of dodecaborides. Their crystal structures, however, have not been studied in such detail and are almost not studied at low temperatures (the ZrB$_{12}$ structure at 140 K \cite{Bol_Leithe-JasperSato02} is a rare exception), although low-temperature physical properties often reveal features that require explanations based on the crystal structure. Moreover, clear explanations are not always easy to obtain in the framework of a simple cubic model. For example, in Refs.~\cite{Bol_CzopnikShitsevalova04, Bol_CzopnikShitsevalova05}, linear thermal expansion coefficients $\alpha$ of $R$B$_{12}$ single crystals, $R$~=~Y, Ho, Er, Tm, Lu, were measured in the temperature range of 5--300 K. The values of the coefficients at low temperatures varied nonlinearly for all the compounds studied. The nonlinear $\alpha(T)$ dependencies had two minima, sometimes negative in magnitude. Two temperature intervals with negative thermal expansion (NTE) were found for LuB$_{12}$ crystals:\index{LuB$_{12}$!negative thermal expansion} the first one was 60--130~K with a minimal negative at 90 K and the second one was 10--20~K with a minimum at 12~K. One NTE interval 50--70~K was revealed for YB$_{12}$ with a negative minimum near 60~K whereas the second minimum at 15~K \mbox{was close to zero but positive}.

The relationship between anomalies of the thermal expansion and crystal structure of the dodecaborides is poorly understood so far. Until recently, published data on the observed anomalies of the $R$B$_{12}$ structure was actually limited to discussions about observed disordering of the metal atoms near the $4a$ position of the $Fm\overline{3}m$ group \cite{Bol_MenushenkovYaroslavtsev13, Bol_WerheitPaderno06} and to a short report on a small tetragonal distortion of the LuB$_{12}$ lattice at temperatures below 150 K \cite{Bol_PietraszkoCzopnik00}, which was discovered in the analysis of thermal expansion using x-ray data. Reports of a tetragonal distortion of the dodecaboride lattice even at room temperature appeared before, but they only concerned ScB$_{12}$ and Sc-containing solid solutions Sc$_{1-x}R_x$B$_{12}$ ($R$~=~Y, Zr) \cite{Bol_HamadaWakata93, Bol_LiangZhang19, Bol_MatkovichEconomy65, Bol_PadernoShitsevalova95}. As noted in \cite{Bol_PadernoShitsevalova95}, the origin of the transformation from cubic to tetragonal structure in ScB$_{12}$ is unclear. The influence of the size of scandium is not obvious, as the radius of scandium is located within the limits of atomic radii of metals that form cubic $R$B$_{12}$ phases. The cell parameters of known dodecaborides are graphically presented in \cite{Bol_PadernoShitsevalova95} as functions of their $d$, 4$f$, or 5$f$ metal radii. All data fit on three straight lines with an individual slope for each group of elements. Yttrium, gadolinium and thorium dodecaborides that finish the corresponding series have almost identical radii but considerably different lattice constants. The $d$-elements form the Hf\,--\,Zr\,--\,Sc\,--\,Y line with intermediate Sc which enters into the composition of the tetragonal ScB$_{12}$ structure. Structural stability and physical properties of $M$B$_{12}$ containing transition-metal elements ($M$~=~Sc, Y, Zr, Hf) were studied \cite{Bol_LiangZhang19} using first-principles calculations supplemented with the x-ray diffraction experiments for ScB$_{12}$ and YB$_{12}$. The tetragonal $I4/mmm$ structure was predicted to be the thermodynamic ground state of ScB$_{12}$ and a metastable state of YB$_{12}$, ZrB$_{12}$, and HfB$_{12}$. Tetragonal ScB$_{12}$ was shown to transform reversibly into the cubic $Fm\overline{3}m$ symmetry group at 510~K, which corresponded to the thermodynamic ground state of YB$_{12}$, ZrB$_{12}$, and HfB$_{12}$ at room temperature. The temperatures of the phase transition between tetragonal and cubic phases of the yttrium and zirconium dodecaborides could be lower than 100~K as predicted in \cite{Bol_LiangZhang19}.\enlargethispage{2pt}

Neutron diffraction measurements have revealed that HoB$_{12}$, TmB$_{12}$ and ErB$_{12}$ have incommensurate magnetic structures \cite{Bol_KohoutBatko04, Bol_SiemensmeyerFlachbart06, Bol_SiemensmeyerHabicht07}. The complex magnetic structure of these materials seems to result from the interplay between the RKKY and dipole-dipole interactions. Strong frustration\index{frustration} of an antiferromagnetic order in the fcc symmetry of the dodecaborides could also play an important~role.

\section{Cooperative Jahn-Teller effect as a driving force behind structural instability in dodecaborides}
\index{Jahn-Teller effect!cooperative|(}

The origin of small structural distortions in $R$B$_{12}$ dodecaborides, which distinctly manifest in the background of an almost unchangeable basic cubic structure of the robust boron network, is a challenging and intriguing problem. In most of the works on rare-earth dodecaborides $R$B$_{12}$, discussion of structural instability is usually based on the two main points: (a) the boron network is rigid and undistorted, (b) the dimension of the B$_{24}$ cavity is oversized with respect to the central metal ion $R$; this results in a rattling character\index{rattling phonon mode} of thermal vibrations of the metal atoms $R$ in the cavities \cite{Bol_SluchankoAzarevich11, Bol_SluchankoAzarevich11a, Bol_SluchankoAzarevich12, Bol_WerheitPaderno06, Bol_WerheitFilipov11}. According to this approach, at low temperatures some fraction of the metal ions shifts from the central position in the B$_{24}$ cages to form a cage-glass state \cite{Bol_SluchankoAzarevich11, Bol_SluchankoAzarevich11a, Bol_SluchankoAzarevich12, Bol_SluchankoBogach18}. These displacements may be caused by lattice defects such as impurities and boron vacancies. These models were extensively used in the analysis of low-temperature structural and magnetic characteristics of $R$B$_{12}$ compounds \cite{Bol_SluchankoAzarevich11, Bol_SluchankoAzarevich11a, Bol_SluchankoAzarevich12, Bol_SluchankoBogach18}.

\begin{figure}[b]
\begin{center}
\includegraphics[width=\textwidth]{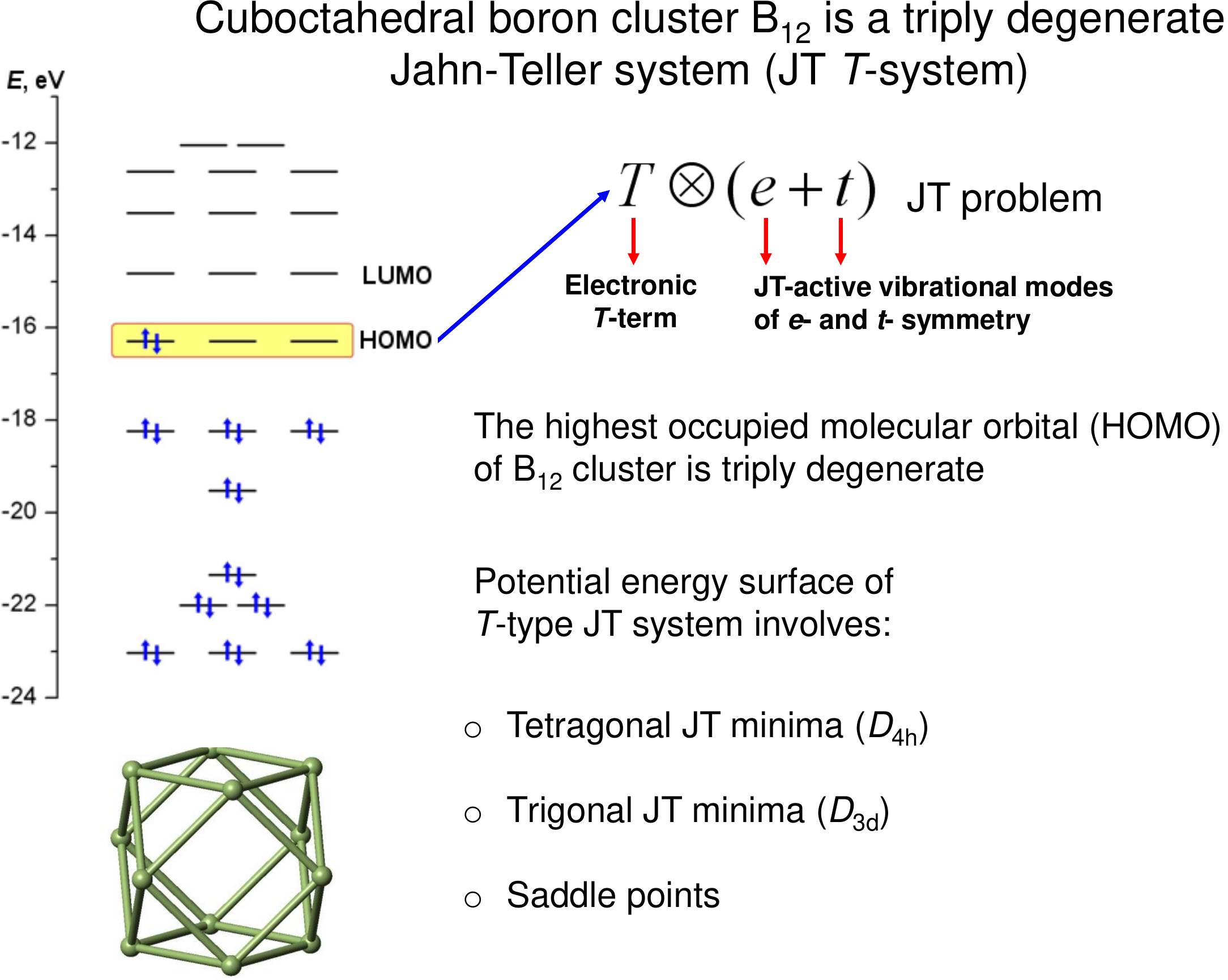}\vspace{-3pt}
\end{center}\index{B$_{12}$ cluster!Jahn-Teller effect}\index{Jahn-Teller effect!in B$_{12}$ clusters}
\caption{On the origin of the Jahn-Teller effect in isolated cuboctahedral B$_{12}$ clusters in $R$B$_{12}$ compounds. The highest occupied molecular orbital (HOMO) is triply degenerate. Partial population of the HOMO orbitals with electrons produces a triply degenerate many-electron ground $T$-state, which leads to Jahn-Teller distortions.}
\label{Fig:Bolotina_JT-effect}
\end{figure}

However, as will be shown below, the basic assumptions of these models need some revision, since they do not take into account some important features of the electronic structure of the $R$B$_{12}$ dodecaborides and their structural units. Namely, apart from displacements of $R$ atoms in the oversized B$_{24}$ cages, low-temperature lattice distortions in $R$B$_{12}$ may originate from intrinsic structural instability of B$_{12}$ cuboctahedra related to the JT effect. This point can be best illustrated for isolated B$_{12}$ units. Similarly to other high-symmetry molecules, such as B$_{12}$ icosahedra in boron and higher metal borides \cite{Bol_FranzWerheit89, Bol_FranzWerheit91}, the cuboctahedral boron clusters B$_{12}$ may have an orbitally degenerate ground state resulting in JT distortions of the regular cubic structure. In terms of molecular orbitals (MOs), the orbital degeneracy of the ground state is associated with partial electron occupation of the highest occupied molecular orbital (HOMO) of B$_{12}$, which is represented by triply degenerate MOs of a $t$-type symmetry (Fig.~\ref{Fig:Bolotina_JT-effect}).\index{B$_{12}$ cluster!occupation of molecular orbitals|(}

\begin{figure}[t]
\begin{center}
\includegraphics[width=\textwidth]{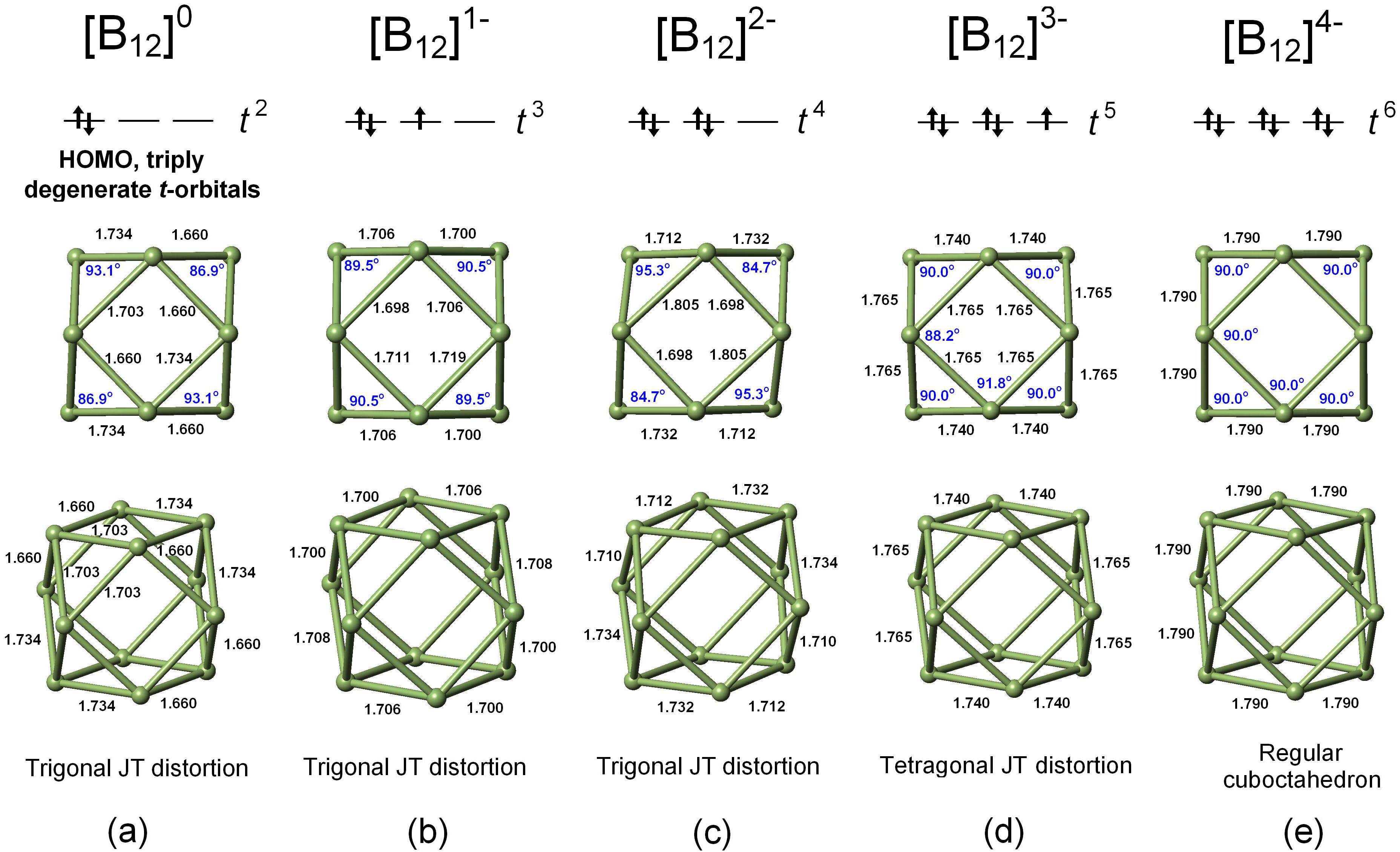}\vspace{-3pt}
\end{center}
\caption{Molecular structure of neutral and negatively charged isolated [B$_{12}$]$^{n-}$ clusters \mbox{($n=0$--4)} obtained from DFT geometry-optimization calculations \cite{Bol_SluchankoBogach18}. The optimized structures correspond to the deeper local JT minima of the corresponding charged cluster. Principal atomic distances (\AA) and bond angles are indicated.}
\label{Fig:Bolotina_B12-clusters}
\end{figure}

Essentially, the character of the JT distortion depends on the electric charge of the B$_{12}$ clusters. In an electrically neutral [B$_{12}$]$^0$ cluster, the HOMO accommodates two electrons ($t^2$ configuration); in negatively charged clusters [B$_{12}$]$^{n-}$ ($n=1$--4), the number $m$ of $t$-electrons in HOMO increases from three to six (Fig.~\ref{Fig:Bolotina_B12-clusters}). Four electronic configurations $t^m$ with $m = 2$--5 produce a triply degenerate many-electron ground $T$-state [Fig.~\ref{Fig:Bolotina_B12-clusters}\,(a--d)] This implies that neutral [B$_{12}$]$^0$ cluster and charged [B$_{12}$]$^{n-}$ clusters ($n = 1, 2, 3$) are JT-active systems, which would tend to distort the cubic structure, except the [B$_{12}$]$^{4-}$ cluster with fully occupied HOMO ($t^6$ configuration), which has a nondegenerate ground state and thus is not JT-active [Fig.~\ref{Fig:Bolotina_B12-clusters}\,(e)]. For JT systems with the $T$-type ground state, the character of distortions is determined by two active vibrational modes of $e$ and $t_2$ symmetry; such situation is referred to as the $T\times(e+t_2)$ Jahn-Teller problem \cite{Bol_BersukerPolinger89}. In this case, depending on the ratio between the strength of the $e$ and $t_2$ electron-vibronic couplings,\index{electron-vibronic coupling} three types of minimum points on the ground-state potential energy surface can occur: trigonal ($D_{3d}$), tetragonal ($D_{4h}$) and orthorhombic ($D_{2h}$) points (Fig.~\ref{Fig:Bolotina_JT-effect}) \cite{Bol_BersukerPolinger89}.

More quantitative information on the amplitude and type of the JT distortions has been obtained from density functional theory (DFT) calculations for the neutral cluster [B$_{12}$]$^0$ and negatively charged [B$_{12}$]$^{n-}$ clusters ($n=1$--4) \cite{Bol_SluchankoBogach18}. Calculated structures of [B$_{12}$]$^{n-}$ ($n = 0$--4) clusters resulting from the DFT geometry optimization are shown in Fig.~\ref{Fig:Bolotina_B12-clusters}. These results indicate that the JT-active clusters [B$_{12}$]$^{n-}$ ($n=0$--3) are slightly distorted cuboctahedra. The character of JT distortions depends strongly on the charge of the cluster: the neutral cluster and charged clusters with $n = 1$,~2 exhibit trigonal type of JT distortions, while the [B$_{12}$]$^{3-}$ cluster shows tetragonal JT distortion. It is important to note that the overall magnitude of the JT distortions in isolated B$_{12}$ clusters is rather small, as the bond lengths and bonding angles vary within $\sim$0.1~\AA\ and $\sim$5$^\circ$, respectively (Fig.~\ref{Fig:Bolotina_B12-clusters}); DFT calculations show that the energy gain resulting from JT distortions is within 0.2--0.3~eV per B$_{12}$ cluster.

These results suggest that the JT structural lability of the B$_{12}$ units in the crystal lattice of metal dodecaborides $R$B$_{12}$ may play an important role in the microscopic mechanism of lattice distortions of $R$B$_{12}$ at low temperature. It should be borne in mind, however, that in the actual crystal structure of $R$B$_{12}$, the B$_{12}$ clusters are connected by B-B covalent bonds to form an extended 3D covalent boron network, in which HOMOs of individual B$_{12}$ clusters may have non-integer electron occupation. Nevertheless, one can expect that in $R$B$_{12}$ crystals some fraction of the JT activity of B$_{12}$ clusters may retain in the three-dimensional boron network because the local triply degenerate HOMOs of B$_{12}$ cuboctahedra remain partially filled, as can be seen from the overall electron balance between the metal and boron sublattices. Due to interactions between the nearest B$_{12}$ clusters in a $R$B$_{12}$ crystal, local JT distortions of B$_{12}$ cuboctahedra become mutually consistent resulting in a symmetry-lowering distortion of the lattice; this phenomenon is known as the cooperative JT effect, which is well documented in the literature \cite{Bol_BersukerPolinger89, Bol_GehringGehring75, Bol_KaplanVekhter95}.\index{B$_{12}$ cluster!occupation of molecular orbitals|)}

In a concentrated JT system, the full JT Hamiltonian of the crystal is given by the equation \cite{Bol_GehringGehring75, Bol_KaplanVekhter95}:
\begin{equation}\label{Bolotina:Eq1}
\hat{H}=\sum_\mathbf{n}\hat{H}_\text{JT}(\mathbf{n})+\frac{1}{2}\hspace{-1ex}\sum_{\substack{\mathbf{n},\mathbf{m}\\(\mathbf{n} \neq \mathbf{m})}}\hspace{-1ex}
\mathbf{Q}^+(\mathbf{n})\hat{K}(\mathbf{n}-\mathbf{m})\mathbf{Q}(\mathbf{m}),
\end{equation}
where the vector indices $\mathbf{n}$ and $\mathbf{m}$ enumerate unit cells of the crystal, $\hat{H}_\text{JT}(\mathbf{n})$ is the one-center JT Hamiltonian for unit cell $\mathbf{n}$, and the last term represents pairwise interactions between the local JT centers $\mathbf{n}$ and $\mathbf{m}$. Here $\mathbf{Q}(\mathbf{n})$ is a vector whose components are the local JT-active vibrational modes and $\hat{K}(\mathbf{n}-\mathbf{m})$ is the operator describing interactions between the local JT vibrational modes on sites $\mathbf{n}$ and $\mathbf{m}$. The electronic and geometric structure of a cooperative JT system is determined from the minimization of the total energy of the crystal resulting from the competition of the local distortions, the first term in Eq.~(\ref{Bolotina:Eq1}), and the interaction between the different sites (the second term). In the general case, cooperative JT interactions can lead to a variety of situations, depending on the specific character of orbitally-degenerate moieties and interplay between the on-site and inter-site JT interactions. In particular, lower symmetry structures in $R$B$_{12}$ resulting from these cooperative interactions can lead to a parallel alignment of all the local distortions of B$_{12}$ cubooctahedra (which is termed as \emph{ferrodistortive} case)\index{ferrodistortive effect} or to a more complicated geometrical arrangement of the local B$_{12}$ distortions (so-called \emph{antiferrodistortive} case).\index{antiferrodistortive phase} In the ferrodistortive phase, the local JT centers are coupled to a strain of the lattice, which changes the shape of the crystal and its unit cell parameters; this strain mode coupling provides an effective long-range interaction between the JT centers.

\begin{figure}[b]
\begin{center}
\includegraphics[width=\textwidth]{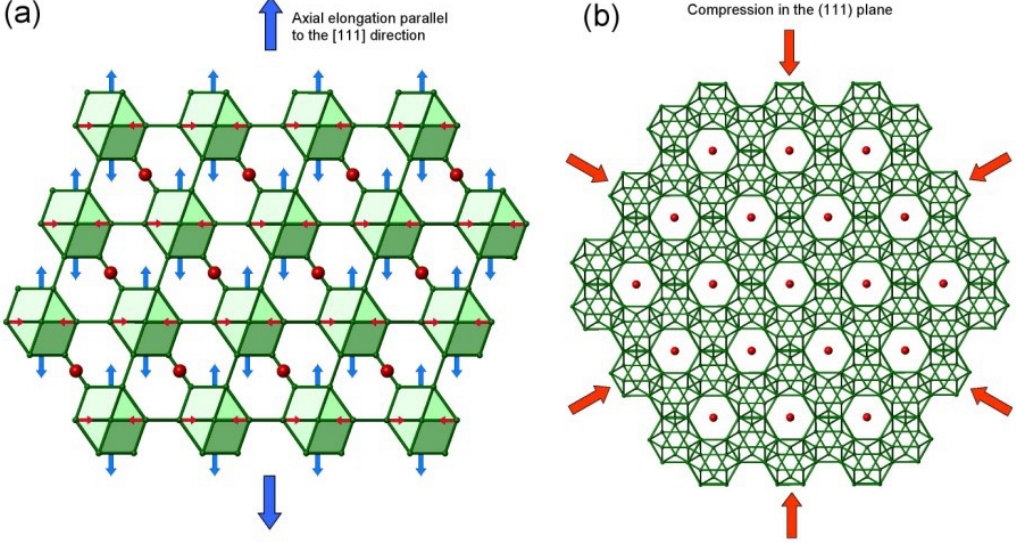}\vspace{-3pt}
\end{center}
\caption{Structure of the ferrodistortive JT phase in $R$B$_{12}$ dodecaborides and the character of the strain of the crystal lattice. (a)~The B$_{12}$ cuboctahedra are all elongated along the trigonal axis [111] and compressed in the orthogonal plane; this geometry corresponds to one of the local trigonal JT minima of B$_{12}$ shown in Fig.~\ref{Fig:Bolotina_B12-clusters}; (b) The crystal lattice of $R$B$_{12}$ is compressed in the (111) plane.}
\label{Fig:Bolotina_RB12-JTphase}
\end{figure}

It is important to note that in most cooperative JT systems the coupling is predominantly to a strain of the lattice \cite{Bol_GehringGehring75, Bol_KaplanVekhter95}. This fact gives an idea of the origin of observed anomalous behavior of $R$B$_{12}$ dodecaborides: in fact, formation of a ferrodistortive JT phase with a long-range ordering of JT distortions mediated by the strain of the lattice is the most likely scenario in metal dodecaborides. More specifically, the ferrodistortive JT phase of $R$B$_{12}$ with the elastic strain axis parallel to the [111] direction seems to be the actual situation that may account for the unusual behavior of $R$B$_{12}$. In this case, when the strain mode corresponds to elongation, the B$_{12}$ cuboctahedra are all elongated along the trigonal axis [111], that refers to one of the trigonal JT minima shown in Fig.~\ref{Fig:Bolotina_B12-clusters}; the local distortion JT axes are all parallel to each other [Fig.~\ref{Fig:Bolotina_RB12-JTphase}\,(a)]. Since the elastic strain does not change the volume of the crystal, elongation in the [111] directions is followed by compression in the (111) plane, as depicted in Fig.~\ref{Fig:Bolotina_RB12-JTphase}\,(b).

\begin{figure}[t]
\begin{center}
\includegraphics[width=0.78\textwidth]{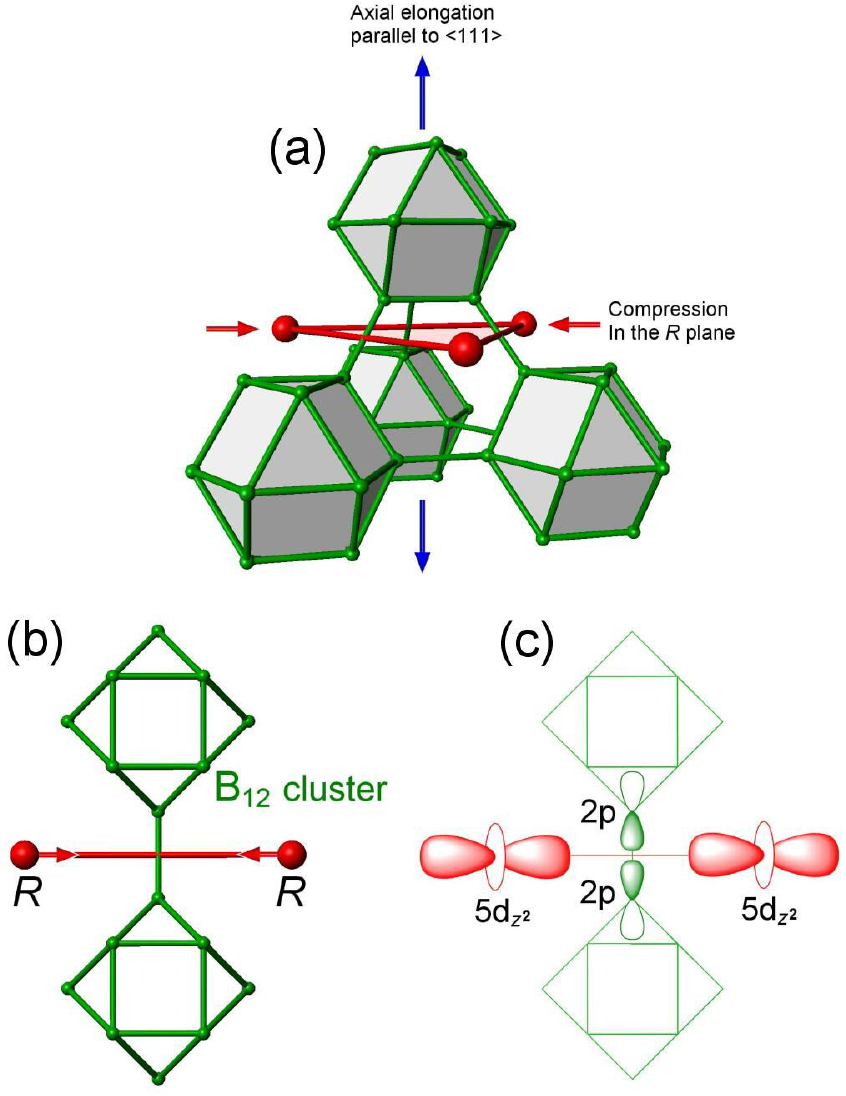}\vspace{-3pt}
\end{center}\index{charge stripes}
\caption{On the origin of charge stripes in the ferrodistortive JT phase in $R$B$_{12}$ dodecaborides, (a)~The general character of the local distortions in the ferrodistortive JT phase, (b) reduction of the $R$-$R$ distance between the neighboring metal atoms due to compression in the (111) plane, (c)~orientation of $5d_{z^2}$ metal orbitals having the strongest $\sigma$-type overlap with the $2p$ valence orbitals of boron; shortening of the $R$-$R$ distance leads to a maximal change in the $5d_{z^2}$($R$)-$2p$(B) orbital overlap, which ultimately causes transfer of excess electron density to the $5d_{z^2}$ orbitals and formation of the charge stripes (see Figs.~\ref{Fig:LuB12-ResidualED} and \ref{Fig:LuB12-MEM-maps} later in the text).}
\label{Fig:Bolotina_RB12-JTdistortions}
\end{figure}

This leads to important changes in the electronic band structure caused by the trigonal strain mode, as the compression in the (111) plane gives rise to some shortening in the $R$-$R$ distance between the neighboring metal atoms [Fig.~\ref{Fig:Bolotina_RB12-JTdistortions}\,(a,b)]. This would change orbital interactions between the $R$ and B atoms responsible for the formation of the electronic conduction band of $R$B$_{12}$, which is mainly represented by $2p$(B) and $5d(R)$ atomic orbitals. The largest changes are expected for the $5d_{z^2}$ metal orbitals that have the strongest $\sigma$-type overlap with the $2p$ valence orbitals of boron [Fig.~\ref{Fig:Bolotina_RB12-JTdistortions}\,(c)]. Accordingly, enhanced $5d_{z^2}$($R$)-$2p$(B) orbital overlap increases the energy dispersion of the electronic conduction band, thereby increasing the overall number of the filled conduction band state below the Fermi level of $R$B$_{12}$. This results in larger electron population of the $5d_{z^2}$($R$) orbitals oriented along the local $R$-$R$ lines connecting neighboring R atoms in the $(111)$ plane [Fig. \ref{Fig:Bolotina_RB12-JTdistortions}\,(c)]. Considering the elongated shape of $5d_{z^2}$($R$) orbitals, this leads to increased electron density along the $R$-$R$ lines being parallel to the side diagonals $[110]$, $[0\overline{1}1]$, $[101]$), which are shown with red solid lines in Fig.~\ref{Fig:Bolotina_RB12-JTstripes}. These findings are in excellent agreement with the recent experimental results on LuB$_{12}$, which reveal lower-symmetry electron density distribution (charge stripes)\index{charge stripes} correlating with the filamentary structure of conduction channels observed in the magnetoresistance measurements \cite{Bol_BolotinaDudka18}. Remarkably, the general character of the residual density distribution near the metal atom in the (100) and (010) planes at low temperatures (50~K) shown in Fig.~\ref{Fig:LuB12-ResidualED} strongly resembles the shape of $5d_{z^2}$($R$) orbitals oriented along the $R$-$R$ lines, as depicted in Fig.~\ref{Fig:Bolotina_RB12-JTdistortions} and Fig.~\ref{Fig:Bolotina_RB12-JTstripes}.

Thus, a theoretical model based on the cooperative JT effect provides a new insight into the microscopic origin of the mysterious structural behavior of $R$B$_{12}$ cubic dodecaborides. The main reason behind the manifestation of low-symmetry effects of $R$B$_{12}$ lies in the inherent structural lability of the B$_{12}$ cuboctahedral units resulting from their orbital degeneracy and the JT effect. Assuming a ferrodistortive JT ordering in $R$B$_{12}$ with deeper trigonal JT minimum of B$_{12}$ clusters enables one to rationalize in a natural way the main experimental results on dodecaborides (see following sections of this chapter for more details):\vspace{-0.6em}
\begin{figure}[t]
\begin{center}
\includegraphics[width=0.75\textwidth]{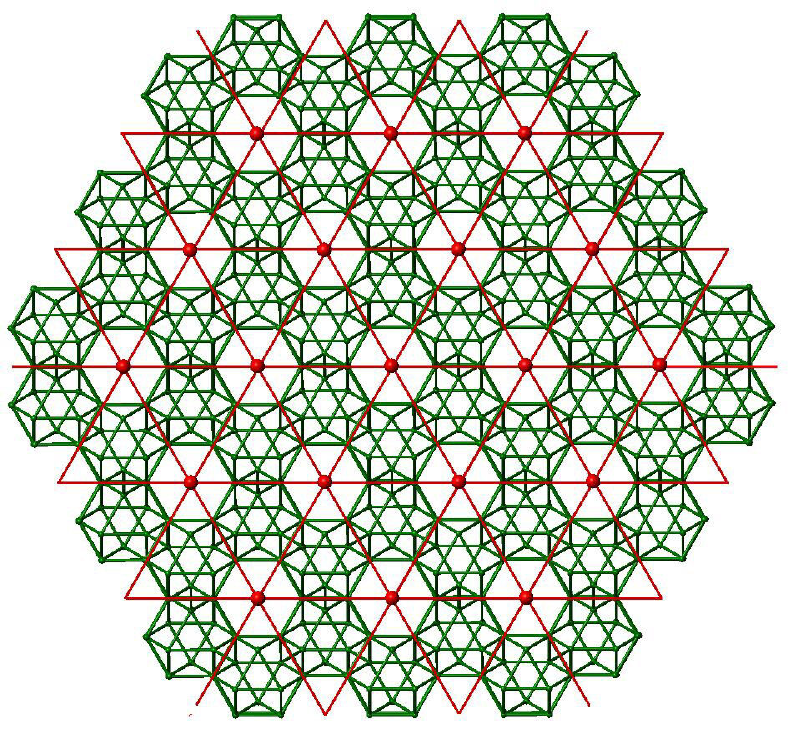}\vspace{-3pt}
\end{center}\index{charge stripes}
\caption{Orientation of the charge stripes (the solid red lines) in the (111) plane of $R$B$_{12}$ resulting from the electron density transfer to the $5d_{z^2}$($R$) orbitals caused by the elastic strain of the crystal lattice in the ferrodistortive JT phase (see Fig.~\ref{Fig:Bolotina_RB12-JTdistortions}).}
\label{Fig:Bolotina_RB12-JTstripes}
\end{figure}
\begin{enumerate}
\item The cooperative JT model explains the presence of small distortions of the cubic lattice of $R$B$_{12}$ at all temperatures, including room temperature. Indeed, in the ferrodistortive JT state distortions persist at all temperatures, without a structural phase transition. Theoretical treatment of the ferrodistortive JT state is similar to that for a magnetic spin lattice in an external magnetic field, which always retains some magnetization \cite{Bol_GehringGehring75, Bol_KaplanVekhter95}; the latter serves as an order parameter, whose analog in $R$B$_{12}$ dodecaborides corresponds to the relative magnitude of the deviation from the regular cubic structure.
\item The JT model provides a physically transparent insight into the origin of lower symmetry electron density distribution in $R$B$_{12}$, including appearance of charge stripes in LuB$_{12}$ and the filamentary structure of conduction channels resulting in anisotropic magnetoresistance \cite{Bol_BolotinaDudka18}. These effects are mainly due to enhanced electron occupation of the $5d_{z^2}$ metal orbitals resulting from the larger $5d(R)$-$2p$(B) orbital overlap caused by elastic shortening in the (111) plane in the ferrodistortive JT state (Figs.~\ref{Fig:Bolotina_RB12-JTphase} and \ref{Fig:Bolotina_RB12-JTdistortions}).
\item The above consideration indicates that low-symmetry distortions are a unique property of all rare-earth dodecaborides, as they result from the JT structural lability of B$_{12}$ units, not from the ground-state characteristics of the metal ions $R$.
\item This theoretical approach evidently demonstrates that subtle structural departures from the regular cubic structure of $R$B$_{12}$ dodecaborides are by no means an artifact of x-ray diffraction analysis, but they are an inherent property of all rare-earth $R$B$_{12}$ compounds resulting from the JT activity of the B$_{12}$ units. Indeed, in the ferrodistortive JT state, arbitrarily weak JT interactions always lead to a static lattice distortion, by analogy with the nonzero magnetization of the spin system in an external magnetic field. More specifically, this property originates from the fact that the energy gain resulting from the JT distortions grows linearly with the strain value $e$, while the competing elastic energy is proportional to $e^2$; therefore, at small strains the low-symmetry JT state lies lower in energy \cite{Bol_BersukerPolinger89, Bol_GehringGehring75, Bol_KaplanVekhter95}.
\item In fact, the presence of significant $^{10}$B/$^{11}$B isotope effects in itself suggests the JT origin of the structural instability, since it documents a breakdown of the Born-Oppenheimer approximation for the orbitally degenerate systems, in which electronic and vibrational motions are no longer independent. Generally, isotope effects are more pronounced for the dynamic JT effect,\index{dynamic Jahn-Teller effect} when the JT stabilization energy competes with the vibrational energy; this situation is likely to occur in $R$B$_{12}$.
\end{enumerate}

\index{Jahn-Teller effect!cooperative|)}

\section{Modeling the dynamics of the dodecaboride lattice using x-ray diffraction data}
\index{rare-earth dodecaborides!lattice dynamics|(}

Experimental conditions, such as the sample temperature, can vary to better identify the barely perceptible symmetry violations, which may appear due to the cooperative JT effect. The most known phenomenon is the temperature dependence of the unit-cell values, which should be monotonous in the absence of a lattice transformation (see below). Additional information can be obtained by analyzing atomic displacement parameters (ADPs) at different temperatures using both experimental and theoretical temperature curves.

The key role in the structure analysis of crystals is assigned to structure factors $F(\mathbf{H})$, which provide a transition from measured intensities of diffraction peaks to the distribution of electron density in the crystal. The expression for $F(\mathbf{H})$ is as follows:
\begin{equation}\label{Bolotina:Eq2}
F(\mathbf{H})=\sum_{\nu=1}^N f_\nu(|\mathbf{H}|)\exp(2\pi i \mathbf{r}_\nu\cdot\mathbf{H})T_\nu(\mathbf{H}).
\end{equation}
Here $\mathbf{H} = h\mathbf{a}^\ast + k\mathbf{b}^\ast + l\mathbf{c}^\ast$ is a scattering vector; $f_\nu(|\mathbf{H}|)$ is the atomic scattering factor of the atom at $\mathbf{r}_\nu$; $T_\nu(\mathbf{H}) = \int\! p(\mathbf{u}_\nu)\exp(2\pi i\mathbf{u}_\nu\!\cdot\!\mathbf{H}){\rm d}^3\mathbf{u}$ is the temperature factor, which is known also as the Debye-Waller factor that accounts for the atomic displacements $\mathbf{u}_\nu$ from the lattice points. Summation is carried out over all atoms in the unit cell of the crystal. As one can see, the temperature factor $T_\nu(\mathbf{H})$ is the Fourier transform of the probability density function $p(\mathbf{u}_\nu)$ whose coefficients are atomic displacement parameters (ADPs) discussed below. The $p(\mathbf{u}_\nu)$ function can be approximated respectively by univariate or trivariate Gaussian in case of isotropic or anisotropic harmonic vibrations of an atom, and it can be more complicated in case of anharmonic vibrations as a result of heating, for instance. In any case, however, the temperature does not participate directly in the calculations either as a fixed parameter or as a refined variable. Moreover, the conventional approach does not require any assumption of the atomic displacement nature. The ADP values may correspond to thermal vibrations supplemented with static shifts \cite{Bol_K.N.TruebloodAbrahams96}. Along with the atomic coordinates, ADPs are the refined parameters of the structural model. The least-squares refinement procedure consists in approximation of $|F_{\rm calc}(\mathbf{H})|^2$ calculated by the formula (\ref{Bolotina:Eq2}) with $|F_{\rm obs}(\mathbf{H})|^2$, whose values are proportional to the measured intensities of the diffraction reflections. The displacements of each atom are represented in the structural model by one or more parameters, depending on the chosen formalism (isotropic, anisotropic harmonic or anharmonic displacements). Harmonic ADPs form a second-rank matrix $\{u_{ij}\}$, $1\leq i, j \leq 3$, the trace of which gives an estimate of the equivalent atomic displacements $\langle u^2 \rangle_{\rm eq}$ or $u_{\rm eq}$ in short notation, $u_{\rm eq}=(u_{11}+u_{22}+u_{33})/3$. This parameter often appears in studies of the thermal properties of solids.

An alternative method of quantifying atomic displacement parameters is not tied directly to a structural model. Thermal vibration amplitudes $u_{\rm calc}(R)$ of the metal atoms in the large cavities of the dodecaboride structure well correspond to the Einstein model \cite{Bol_Einstein07} for independent harmonic oscillators supplemented with a temperature independent static component $\langle u^2 \rangle_{\rm shift}$ or $u_{\rm shift}$ in short notation:
\begin{equation}\label{Bolotina:Eq3}
u_{\rm calc}(R)=\frac{\hbar^2}{k_{\rm B}m_{\rm a}T_{\rm E}}\left(\frac{1}{2}+\frac{1}{\exp(T_{\rm E}/T-1)}\right)+u_{\rm shift}(R)
\end{equation}
The expanded Debye model \cite{Bol_Debye14} is suitable for atoms of the boron framework whose displacements strongly correlate with each other:
\begin{equation}\label{Bolotina:Eq4}
u_{\rm calc}(\text{B})=\frac{3\hbar^2}{k_{\rm B}m_{\rm a}T_{\rm D}}\left(\frac{1}{4}+\left(\frac{T}{T_{\rm D}}\right)^{\!2}\hspace{-3pt}\int_0^\frac{T_{\rm D}}{T^{\phantom{!}}}\hspace{-3pt}\frac{y\,{\rm d}y}{\exp(y)-1}\right)+u_{\rm shift}(\text{B})
\end{equation}
The agreed notations are: $\hslash=h/2\pi$ is the Planck constant; $k_{\rm B}$\,---\,the Boltzmann constant; $m_{\rm a}$\,---\,atomic mass; $T_{\rm E}$ ($T_{\rm D}$)\,---\,the characteristic Einstein (Debye) temperature; $T$\,---\,the temperature of the experiment.

The problem is that the characteristic Einstein (Debye) temperature and the value of $u_{\rm shift}$ must be known in advance to calculate the values of $u_{\rm calc}$ from Eqs.~(\ref{Bolotina:Eq3}) or (\ref{Bolotina:Eq4}). Still, it is possible to solve the inverse problem of calculating the characteristic Einstein (Debye) temperature and $u_{\rm shift}$ using the values of $u_{\rm eq}$ determined from diffraction data. For this purpose, one should collect the multi-temperature data sets $\{h, k, l, |F_{\rm obs}|, \sigma_F\}$ and refine the crystal structure at each temperature using conventional techniques. As a result, each atom is supplied at each temperature with a value of $u_{\rm obs} = u_{\rm eq}$. The multi-temperature set of these parameters then serves as an input for a least-squares procedure $\sum|u^2_{\rm calc} - u^2_{\rm obs}|\rightarrow \min$ to fit the model curve to the set of $u_{\rm obs}$. The characteristic Einstein (Debye) temperature $T_{\rm D}$ ($T_{\rm E}$) and the value of $u_{\rm shift}$ are adjustable parameters of this procedure \cite{Bol_DudkaBolotina19}.

\begin{figure}[b!]
\begin{center}
\includegraphics[width=0.96\textwidth]{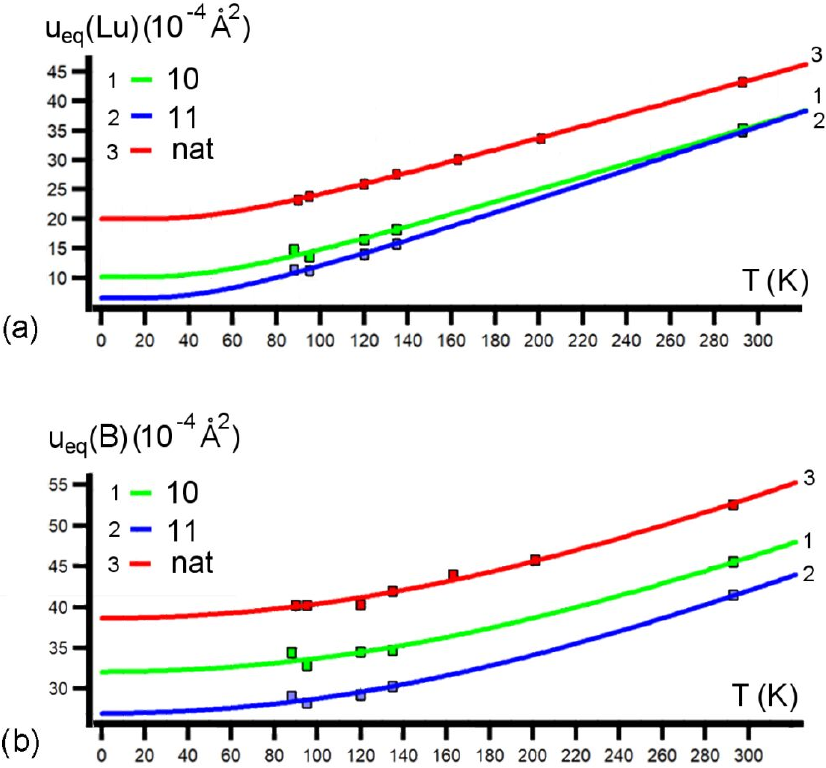}\vspace{-3pt}
\end{center}
\caption{Temperature dependencies of $u_{\rm eq}$ in the crystals of Lu$^N$B$_{12}$ ($N = 10$, 11, nat). The Einstein (a) and Debye (b) models are used respectively for Lu and B atoms. The fit is based on the $u_{\rm obs}$ values marked with squares \cite{Bol_BolotinaDudka19}.}
\label{Fig:Bolotina7}
\end{figure}

\begin{figure}[b]
\begin{center}
\includegraphics[width=\textwidth]{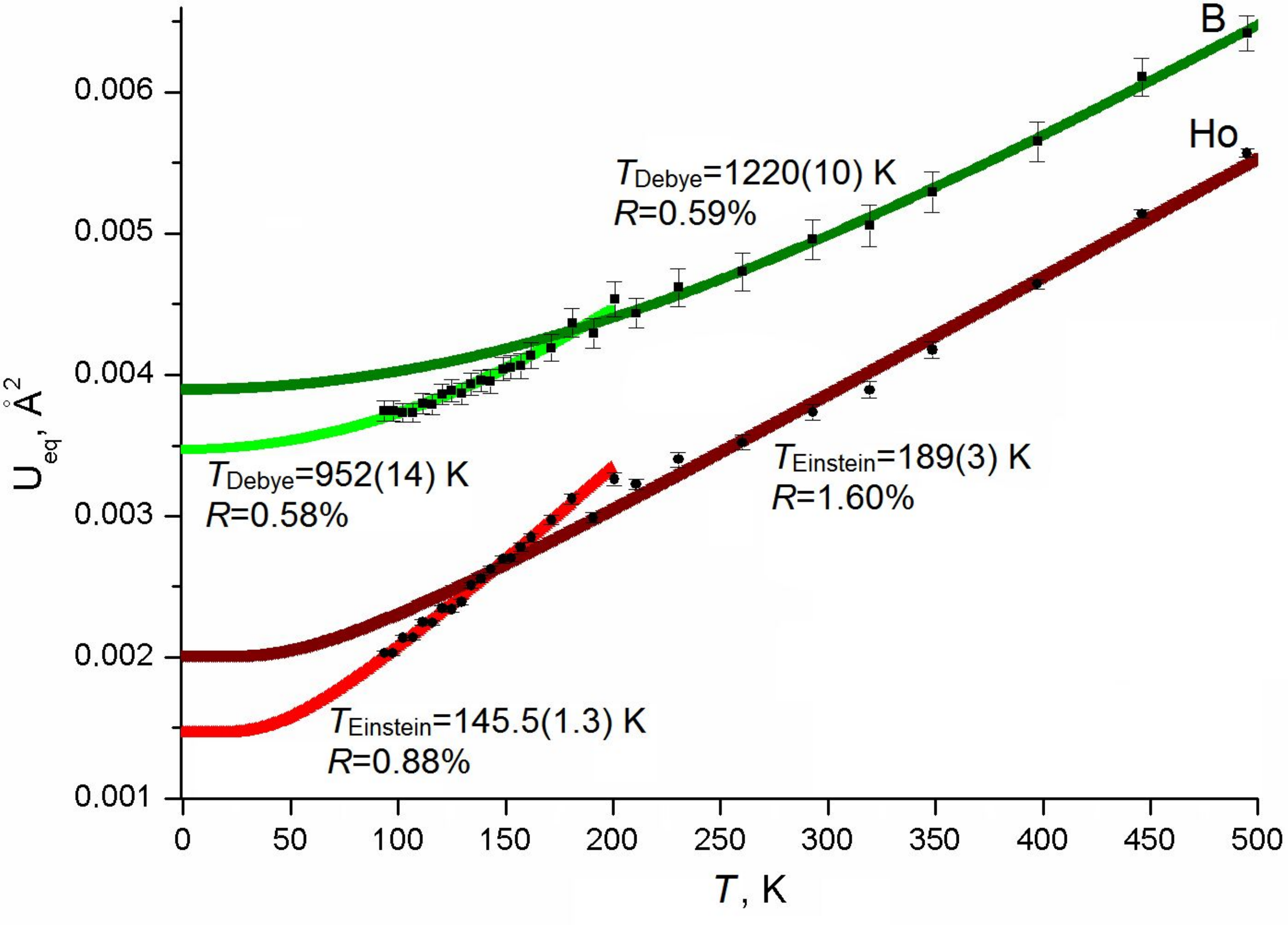}\vspace{-3pt}
\end{center}
\caption{Experimental ADPs ($u_{\rm eq}$) in HoB$_{12}$ are fitted using the Einstein (for Ho) and Debye (for B) models in temperature ranges 86--180 and 210--500~K. $R = \sum |u^2_{\rm obs} - u^2_{\rm calc}|/\sum u^2_{\rm obs}$.}
\label{Fig:Bolotina8}
\end{figure}

Thus, the proposed approach \cite{Bol_DudkaBolotina19} allows us to solve several problems at once:\vspace{-0.6em}
\begin{enumerate}
\item to obtain the Einstein (Debye) characteristic temperatures estimated otherwise from the heat capacity or somehow else;
\item to describe the temperature dependence of the thermal atomic vibrations using an appropriate analytical function;
\item to separate the contributions of the static and dynamic components into the equivalent parameter $u_{\rm eq}$ of atomic displacements.
\end{enumerate}\vspace{-0.6em}
The Debye and Einstein models were previously used to fit the multi-temperature ADPs and to estimate the Debye (Einstein) temperatures in crystals of various compositions including hexaborides $R$B$_6$ ($R$~=~Y, La\,--\,Gd) \cite{Bol_Korsukova94, Bol_TakahashiOhshima99, Bol_VATrounovChernyshev93, Bol_TrounovMalyshev94}. This approach, being first applied to the dodecaborides Lu$^N$B$_{12}$ ($N = 10$, 11, nat), revealed a difference in the static components $u_{\rm shift}$ depending on the isotope composition of boron \cite{Bol_BolotinaDudka19}. The abbreviation 'nat' is hereinafter used to refer to natural boron with the ratio $^{10}$B\,:$^{11}$B~$\approx$~19.8\,:\,80.2. Refined values of $T_{\rm E}$ ($T_{\rm D}$) and $u_{\rm shift}$ were substituted in the Eqs.~(\ref{Bolotina:Eq3}) and (\ref{Bolotina:Eq4}) to draw the curves for Lu and B presented in Figs.~\ref{Fig:Bolotina7}\,(a) and (b), respectively.

The ADPs sum up mean-square zero vibrations $\langle u^2\rangle_{\rm zero}$, temperature-dependent thermal vibrations $\langle u^2(T)\rangle$, and static shifts $\langle u^2\rangle_{\rm shift}$. The curves in Fig.~\ref{Fig:Bolotina7} are plotted for three crystals with very close Debye (Einstein) temperatures, so that $u_{\rm eq}(T)$ mostly differ in their temperature independent components $\langle u^2\rangle_{\rm c}=\langle u^2\rangle_{\rm zero}+\langle u^2\rangle_{\rm shift}$. As shown in \cite{Bol_BolotinaDudka19}, these components are maximal in Lu$^{\rm nat}$B$_{12}$ both for Lu and B atoms. Static distortions of boron polyhedra are combined with static shifts of Lu atoms from the lattice points, which can be explained by disorder in $^{10}$B\,-$^{11}$B substitution in the crystal with natural boron.

The structure of single-crystal HoB$_{12}$ was studied by x-ray diffraction analysis in the $Fm\overline{3}m$ group at 29 temperatures in the range of 86--500~K \cite{Bol_Khrykina}. Temperature variations of $u_{\rm eq}(\text{B})$ and $u_{\rm eq}(\text{Ho})$ lose stability near 200~K. To improve the fit, one has to divide each of the experimental sets of $u_{\rm eq}$ into two parts obtained in the 86--180~K and 210--500~K temperature ranges, and to build two curves for each atom (Fig.~\ref{Fig:Bolotina8}) for better modeling of the experimental curves. The instability of the unit-cell values could be clearly determined from the x-ray data not only in HoB$_{12}$ in the temperature range 150--200~K, but also in $R$B$_{12}$ ($R$~=~Ho, Tm, Yb, Lu) below 200~K (see Figs.~\ref{Fig:Bolotina9}--\ref{Fig:Bolotina13-14} in the next section). The development of similar lattice instability with decreasing temperature was also reported earlier in Lu$^{N}$B$_{12}$ crystals with different isotopic boron composition ($N$~=~10, 11, nat) that were studied using low-temperature heat capacity and Raman scattering data \cite{Bol_SluchankoAzarevich11}. The maximum density of vibrational states was observed at the temperature near 150~K \cite{Bol_SluchankoAzarevich11}. It was noted that the mean free path of phonons reaches the Ioffe-Regel limit in the vicinity of this temperature, being compared with their wavelength. Remarkable spectral changes in the zero-field spectra and a sharp maximum in the relaxation rate were recorded near 150~K in $\mu$SR experiments for dodecaborides $R$B$_{12}$ ($R$~=~Yb, Lu) and solid solutions Lu$_{1-x}$Yb$_x$B$_{12}$. It has been suggested that the large-amplitude dynamic features arise from atomic motions within the B$_{12}$ clusters \cite{Bol_KalviusNoakes02, Bol_KalviusNoakes03}. Most likely, the instability of $u_{\rm eq}$ is caused by changes in the phonon structure of the rare-earth dodecaborides and is not a unique feature of only HoB$_{12}$.
\index{rare-earth dodecaborides!lattice dynamics|)}

\section{Crystal structure: problems and results}

\subsection{The Jahn-Teller distortions of structural parameters}
\index{rare-earth dodecaborides!Jahn-Teller distortions|(}

\begin{figure}[!t]
\begin{center}
\includegraphics[width=\textwidth]{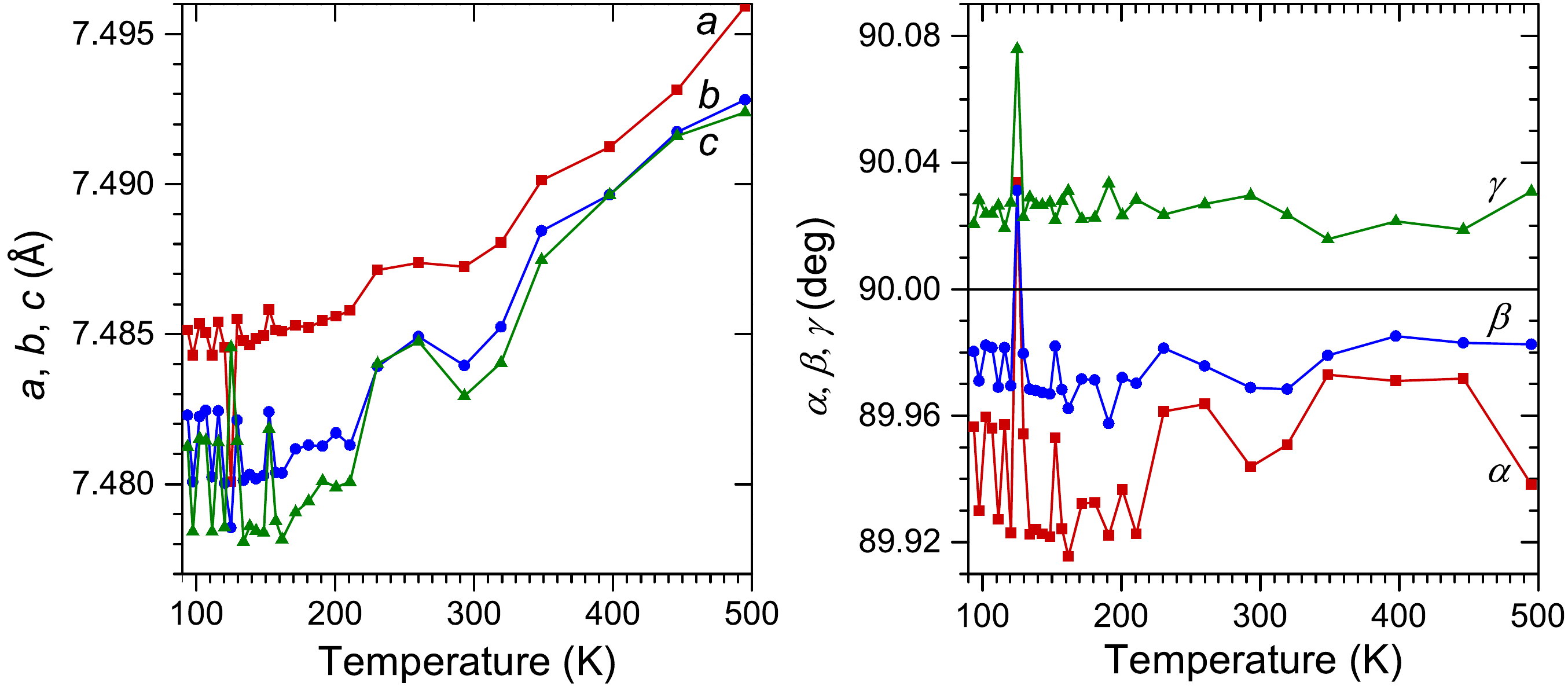}\vspace{-12pt}
\end{center}
\caption{Linear ($a$, $b$, $c$) and angular ($\alpha$, $\beta$, $\gamma$) unit-cell parameters of HoB$_{12}$ in the temperature range 85--500~K \cite{Bol_Khrykina}.\vspace{3pt}}
\label{Fig:Bolotina9}
\begin{center}
\includegraphics[width=\textwidth]{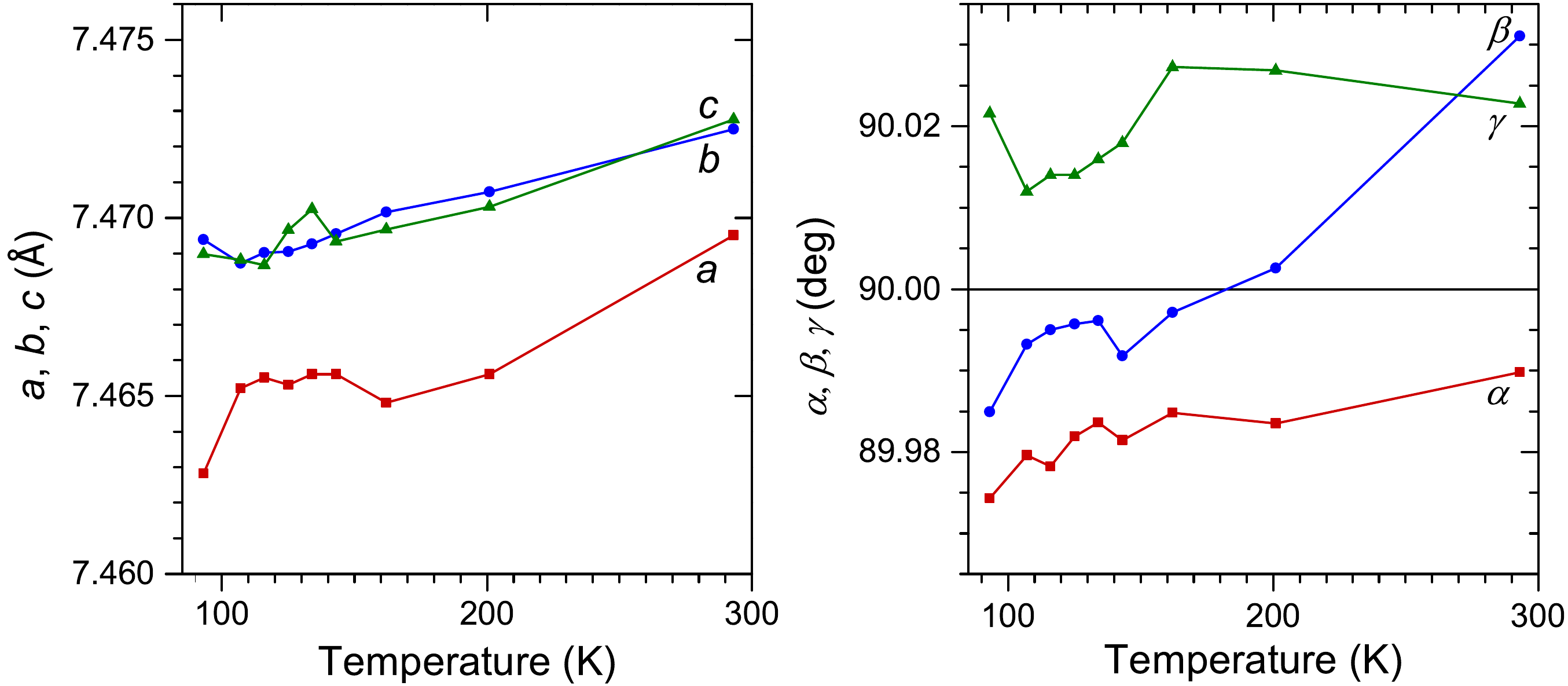}\vspace{-12pt}
\end{center}
\caption{Linear ($a$, $b$, $c$) and angular ($\alpha$, $\beta$, $\gamma$) unit-cell parameters of TmB$_{12}$ in the temperature range 85--300~K \cite{Bol_DudkaKhrykina19}.\vspace{3pt}}
\label{Fig:Bolotina10}
\begin{center}
\includegraphics[width=\textwidth]{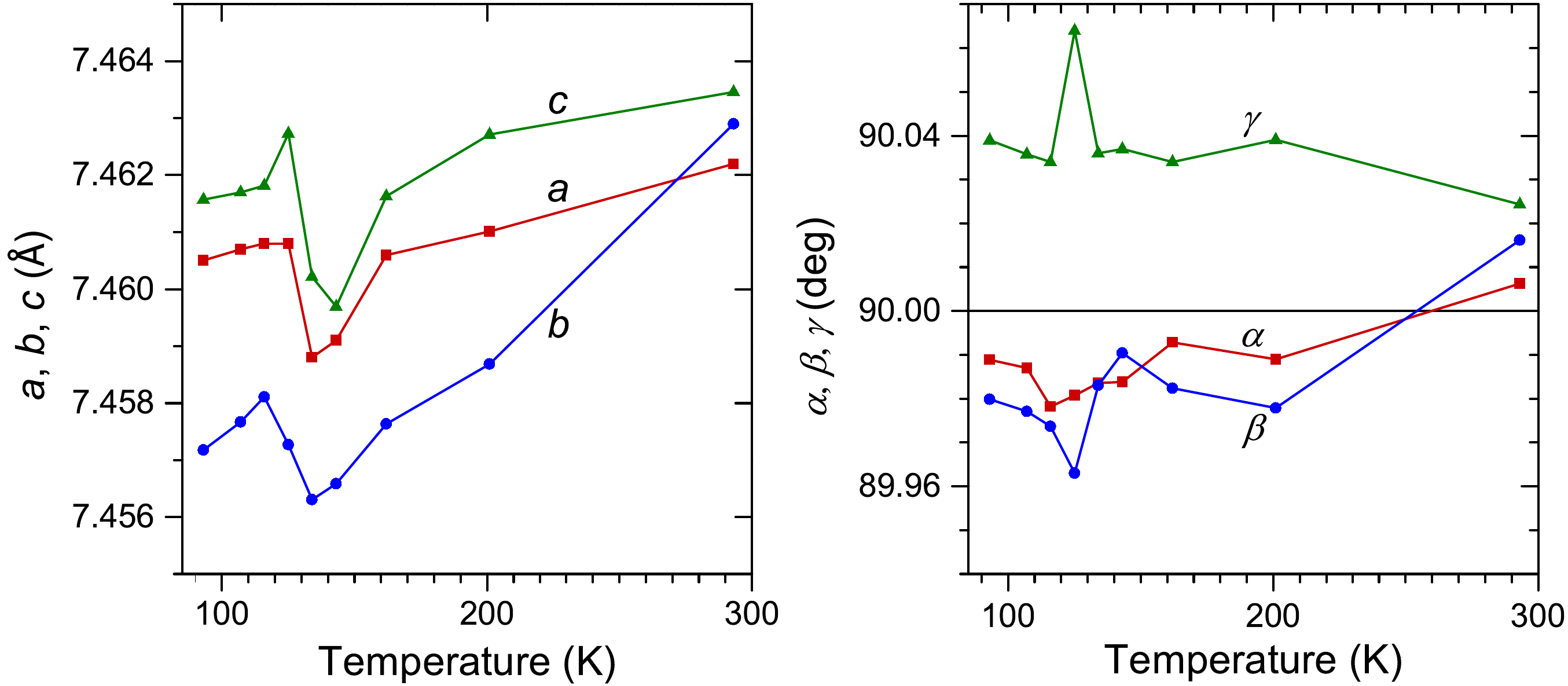}\vspace{-12pt}
\end{center}
\caption{Linear ($a$, $b$, $c$) and angular ($\alpha$, $\beta$, $\gamma$) unit-cell parameters of YbB$_{12}$ in the temperature range 85--300~K \cite{Bol_DudkaKhrykina}.}
\label{Fig:Bolotina11}\vspace{-2em}
\end{figure}

Active studies of the $R$B$_{12}$ structure at various temperatures, which were started in the early 2000s but not continued at that time, were resumed later after the tetragonal distortion of the LuB$_{12}$ structure had been confirmed in the temperature range 50--75~K \cite{Bol_BolotinaVerin16}. The structure of a LuB$_{12}$ single crystal was then thoroughly studied at room temperature \cite{Bol_DudkaKhrykina17}. The single crystals of LuB$_{12}$ were grown by modified crucibleless inductive floating zone melting using high-purity source materials: lutetium oxide Lu$_2$O$_3$ and boron \cite{Bol_WerheitPaderno06}. One of the purposes of the re-examination of the known structure was to assess the suitability of the grown single crystals for accurate structure analysis. The refinement of the structural model in the $Fm\overline{3}m$ symmetry group with a uniquely low residual factor $R = 0.2$\% was made possible due to the high diffraction quality of the single crystals combined with a set of original experimental techniques \cite{Bol_Dudka07, Bol_Dudka10, Bol_Dudka18} that ensured the accuracy and reliability of the x-ray data measured.

Besides that, accurate measurements of the periods of the LuB$_{12}$ crystal lattice were carried out in the temperature range 20--295~K \cite{Bol_DudkaSmirnova17}. Two periods \mbox{$a \approx b$} did not differ within the limits of the standard uncertainty ($\sigma$), but the third period $c$ steadily deviated downward by $2\sigma$ or more over practically the entire temperature range. In absolute values, the difference in the lattice constants is very small (about 0.002~\AA), which is an order of magnitude less than in the lattice of ScB$_{12}$. Such a small difference in the lattice constants does not give grounds for a revision of the structural model, especially in view of what was said above about the excellent results of the refinement of the cubic structure of LuB$_{12}$. However, even very small differences in the lattice constants can have a significant effect on the physical properties of crystals. The lattice parameters must be determined for many dodecaborides of different composition in a wide temperature range without symmetry restrictions in order to collect experimental information on the Jahn-Teller distortions. This work is still far from complete, both in the number of crystals studied and in the number of temperature points measured. After the first experiments with LuB$_{12}$,\index{LuB$_{12}$!unit-cell parameters} linear and angular unit-cell parameters have been determined at various temperatures for HoB$_{12}$,\index{HoB$_{12}$!unit-cell parameters} TmB$_{12}$,\index{TmB$_{12}$!unit-cell parameters} and YbB$_{12}$\index{YbB$_{12}$!unit-cell parameters} as shown in Figs.~\ref{Fig:Bolotina9}\,--\,\ref{Fig:Bolotina11}.

The linear parameters always manifest small tetragonal-type distortions. Note that one lattice constant of LuB$_{12}$ and TmB$_{12}$ is smaller than the other two: \mbox{$a \approx b > c$} (the unit cell is slightly compressed along an edge), whereas one lattice constant of HoB$_{12}$ is slightly elongated: \mbox{$a > b \approx c$}. Both linear and angular parameters of each unit cell undergo the most noticeable non-linear changes in the same low-temperature region between 100 and 150~K. It is noteworthy that lattices of TmB$_{12}$ and YbB$_{12}$ undergo opposite changes despite the proximity of Tm and Yb in the series of rare earth elements. Closer to the middle of the mentioned temperature range, the lattice constants of YbB$_{12}$ abruptly decrease and return to the former, even slightly larger values with a further decrease in temperature. The obliquity of the YbB$_{12}$ lattice slightly increases, but then the angles return to their previous values. At the same temperatures, the periods of the TmB$_{12}$ lattice slightly increase, and the angles become slightly closer to 90$^\circ$.
\index{rare-earth dodecaborides!Jahn-Teller distortions|)}

\subsection{Structural peculiarities of dodecaborides different in isotopic boron composition}
\index{rare-earth dodecaborides!crystal structure!isotope dependence|(}

\begin{figure}[t]
\begin{center}
\includegraphics[width=0.75\textwidth]{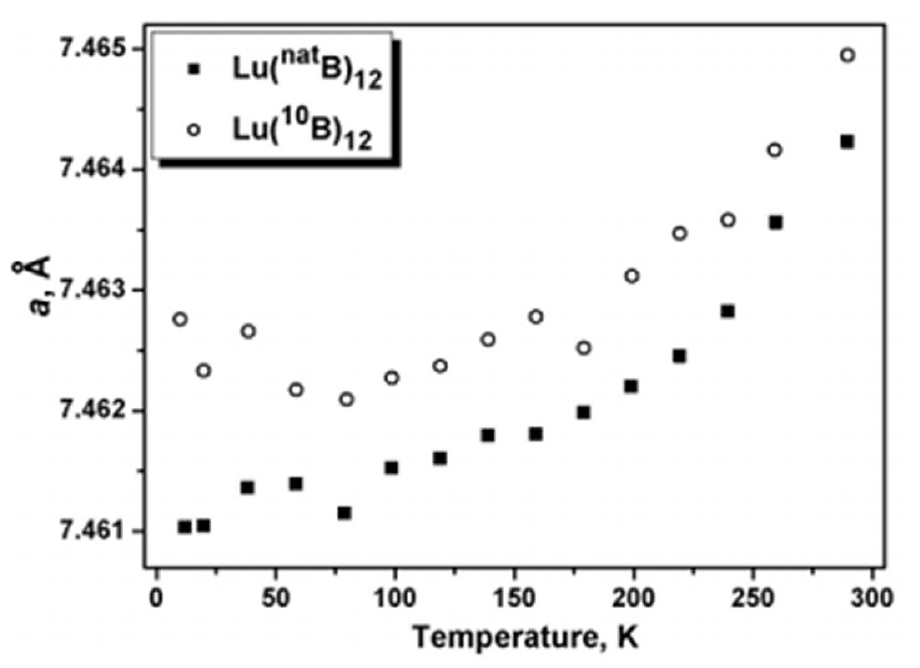}\vspace{-3pt}
\end{center}
\caption{The lattice parameter vs. temperature in LuB$_{12}$ containing natural boron and $^{10}$B isotope~\cite{Bol_Mori08}.}
\label{Fig:LuB12-lattice}
\end{figure}

Since the cooperative JT effect is determined by the dynamics of light boron atoms, one can suppose that isotope substitutions $^{10}$B\,-$^{11}$B may affect both properties and crystal structure of dodecaborides. Till recently, the research was mainly limited to physical properties \cite{Bol_SluchankoAzarevich11, Bol_SluchankoAzarevich11a, Bol_SluchankoAzarevich12, Bol_WerheitPaderno06, Bol_WerheitFilipov11}. Thermal expansion of Lu$^{10}$B$_{12}$ and Lu$^\text{nat}$B$_{12}$ was studied based on the x-ray powder diffraction data in the temperature range 10--290~K~\cite{Bol_MoriGumeniuk07}. Both samples showed negative thermal expansion between 50 and 100~K (Fig.~\ref{Fig:LuB12-lattice}).

This is consistent with the temperature region in which the negative thermal expansion was previously observed for Lu$^\text{nat}$B$_{12}$ by the three-terminal capacitive method \cite{Bol_CzopnikShitsevalova05}. The lattice constant of Lu$^{10}$B$_{12}$ is increased relative to that of Lu$^\text{nat}$B$_{12}$ by 0.001--0.002~\AA\ over the measured temperature range. The $\beta$-rhombohedral boron lattices have the same property, but the difference between the lattice parameters in the crystals with 10\% and 97\% content of $^{10}$B is more noticeable (about 0.03~\AA) as established in Ref.~\cite{Bol_GabuniaTsagareishvili09} whose authors presented a theoretical justification for such an expansion of the $^{10}$B lattice.

\begin{figure}[!t]
\begin{center}
\includegraphics[width=\textwidth]{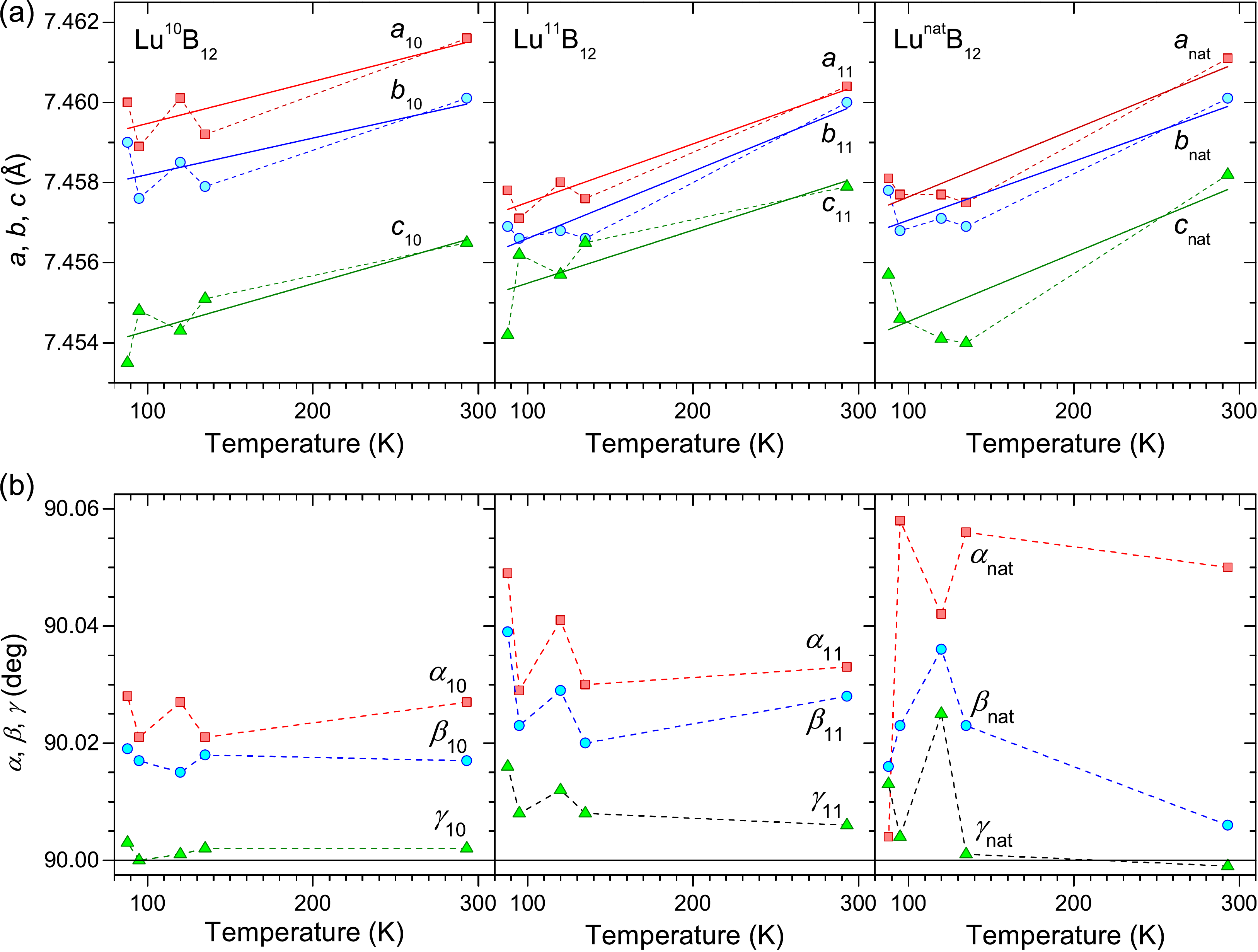}\vspace{-5pt}
\end{center}
\caption{Temperature dependences of the lattice parameters for Lu$^N$B$_{12}$ over the temperature range 88--293~K: (a)~lattice constants; (b)~unit-cell angles. Experimental values are connected by dashed lines; solid lines in panel (a) are linear fits. Standard uncertainties do not exceed 0.0002~\AA\ and 0.001$^\circ$, respectively \cite{Bol_BolotinaDudka19}.}
\label{Fig:Bolotina13-14}
\end{figure}

An influence of the isotopic composition on the structure and properties of Lu$^N$B$_{12}$, $N = 10$,~11,~nat, was studied in \cite{Bol_BolotinaDudka19}. Taking into account both linear and angular distortions of the unit cells of the three crystals, one can conclude that the Lu$^{10}$B$_{12}$ lattice is distorted rather by tetragonal type $a \approx b > c$ whereas the distortions of the Lu$^{11}$B$_{12}$ lattice are more similar to pseudo-trigonal ones $a \approx b \approx c$, $\alpha \approx \beta \approx \gamma > 90^\circ$. Lattice distortions of Lu$^\text{nat}$B$_{12}$ have an intermediate character (see Fig.~\ref{Fig:Bolotina13-14}).

At temperatures below 140~K, the distortions are nonlinear, as can be assumed despite the small number of the points measured. Nonlinear distortions of the parameters, which occur at close temperatures in three different crystals, are hardly explained by a sheer accident. At a temperature of about 120~K, the trigonal-type distortions of the lattices of Lu$^{11}$B$_{12}$ and Lu$^\text{nat}$B$_{12}$ are amplified, as well as the pseudo-tetragonal lattice distortions of Lu$^{10}$B$_{12}$ crystal, but the situation changes again with a further decrease in temperature. Thus, we observe the same jump in the parameters of the unit cell approximately in the middle of the temperature range 100--150~K as in other three dodecaborides mentioned above. It is worth noting that 120~K is close to the upper boundary of the temperature interval with negative thermal expansion of Lu$^\text{nat}$B$_{12}$ according to Refs.~\cite{Bol_CzopnikShitsevalova04, Bol_CzopnikShitsevalova05}.
\index{rare-earth dodecaborides!crystal structure!isotope dependence|)}

\subsection{Formation of charge stripes in voids of the crystal lattice}\index{charge stripes|(}

The numerical differences between the lattice parameters are very small and do not require a transition to the low-symmetry structure model. The crystal structures of LuB$_{12}$, TmB$_{12}$, HoB$_{12}$ and many other dodecaborides can be successfully refined in the cubic group $Fm\overline{3}m$ with low values of $R$-factors. It should be noted, however, that the completeness of the structural analysis is judged not only by the $R$-factor value but also by the distribution of the residual electron density (ED) on the difference Fourier maps. Fourier synthesis of the electron density is a computational procedure, which starts with a set of both experimental and previously calculated parameters. The computational formula can be written in general terms as follows:
\begin{equation}\label{Bolotina:Eq5}
G(\mathbf{r})=\frac{1}{V}\sum_\mathbf{H} A(\mathbf{H})\exp[i\varphi(\mathbf{H})]\exp(-2\pi i\,\mathbf{H}\!\cdot\!\mathbf{r}).
\end{equation}
Here $G(\mathbf{r})$ is either full ($g$) or residual ($\Delta g$) electron density, resulting respectively either from a ``regular'' or difference Fourier synthesis; $V$ is the unit-cell volume; $\mathbf{H} = \sum_i^{\phantom{\ast}}\! h_i^{\phantom{\ast}}\!\mathbf{a}_i^\ast$ is a scattering vector; and $\varphi(H)$ is a scattering phase. $A(\mathbf{H})$ are coefficients dependent on the type of the Fourier synthesis. In case of difference Fourier synthesis, $A(\mathbf{H}) = \bigl||F_{\rm obs}(\mathbf{H})| - |F_{\rm calc}(\mathbf{H})|\bigr|$ is a difference between observed and calculated absolute values of the structure factor. The first value is the square root of the reflection intensity whereas the second one is calculated from atomic coordinates and ADPs, whose values are refined using a least-square technique.

As follows from Eq.~(\ref{Bolotina:Eq5}), the Fourier synthesis of the electron density does not require any data on the crystal symmetry. It can be performed independently in each point of the crystal lattice. Nevertheless, the symmetry of the crystal is usually taken into consideration in the algorithms that implement Fourier synthesis of the electron density. It means that the measured intensities of x-ray reflections are averaged in the corresponding Laue class and the Fourier synthesis is performed in a symmetrically independent region of the unit cell. As a result, the symmetry of the Fourier map exactly corresponds to the space group, information about which is fed to the input of the computational procedure. Certainly, any measurement is not free from the influence of instrumental errors and the data processing methods. On the one hand, the above-mentioned techniques of calculations are designed to improve the accuracy of the results and to ensure visual consistency of the Fourier maps with the stated symmetry of the crystals. On the other hand, averaging can harm since the symmetry of the electron-density distribution over the cell can be overestimated.

\begin{figure}[t!]
\includegraphics[width=\textwidth]{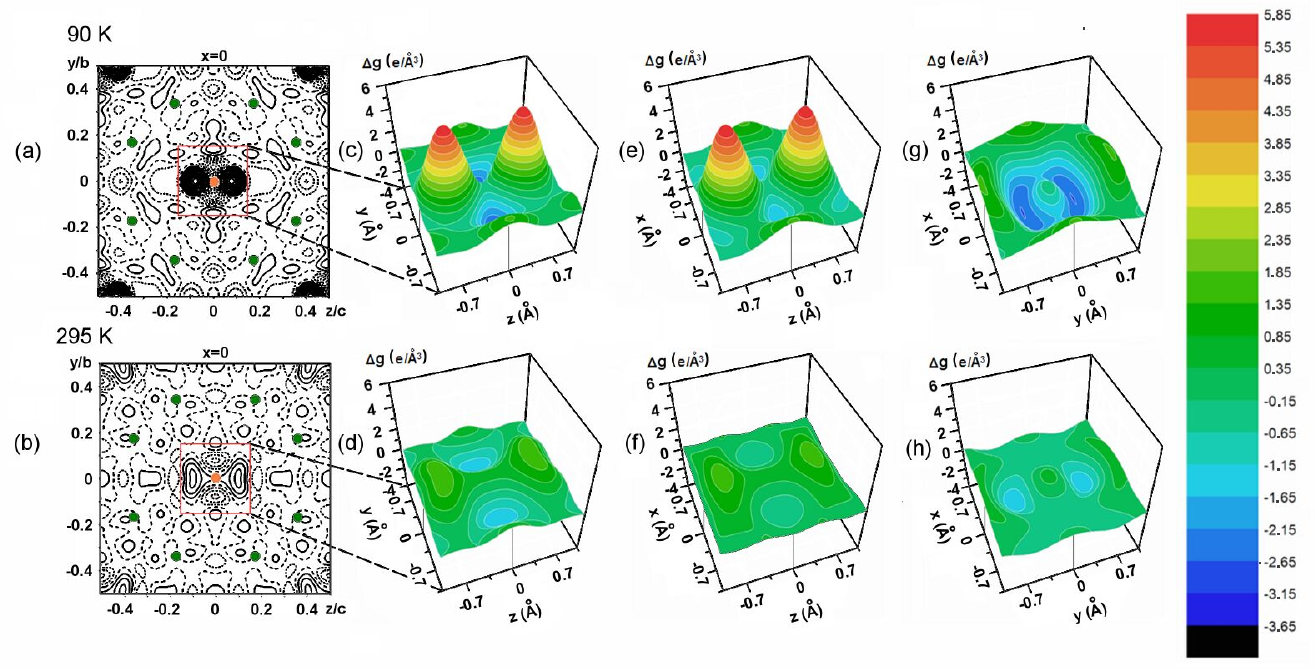}
\index{LuB$_{12}$!electron density}
\caption{Difference Fourier maps (residual electron density $\Delta g$ in $e$/\AA$^3$) in the $x = 0$ face of the LuB$_{12}$ unit cell at (a)~90~K and (b)~295~K. Red circle is the Lu site; green circles are B sites. The panels (c)\,--\,(g) and (d)\,--\,(e) are the surface plots of difference Fourier maps in the vicinity of the Lu ion, in the $x = 0$, $y = 0$, and $z = 0$ faces of the unit cell, respectively. The first and second rows of the figure correspond to temperatures 90~K and 295~K, respectively~\cite{Bol_SluchankoBogach18}. }
\label{Fig:LuB12-DiffFourier}
\end{figure}

\begin{figure}[t!]
\includegraphics[width=\textwidth]{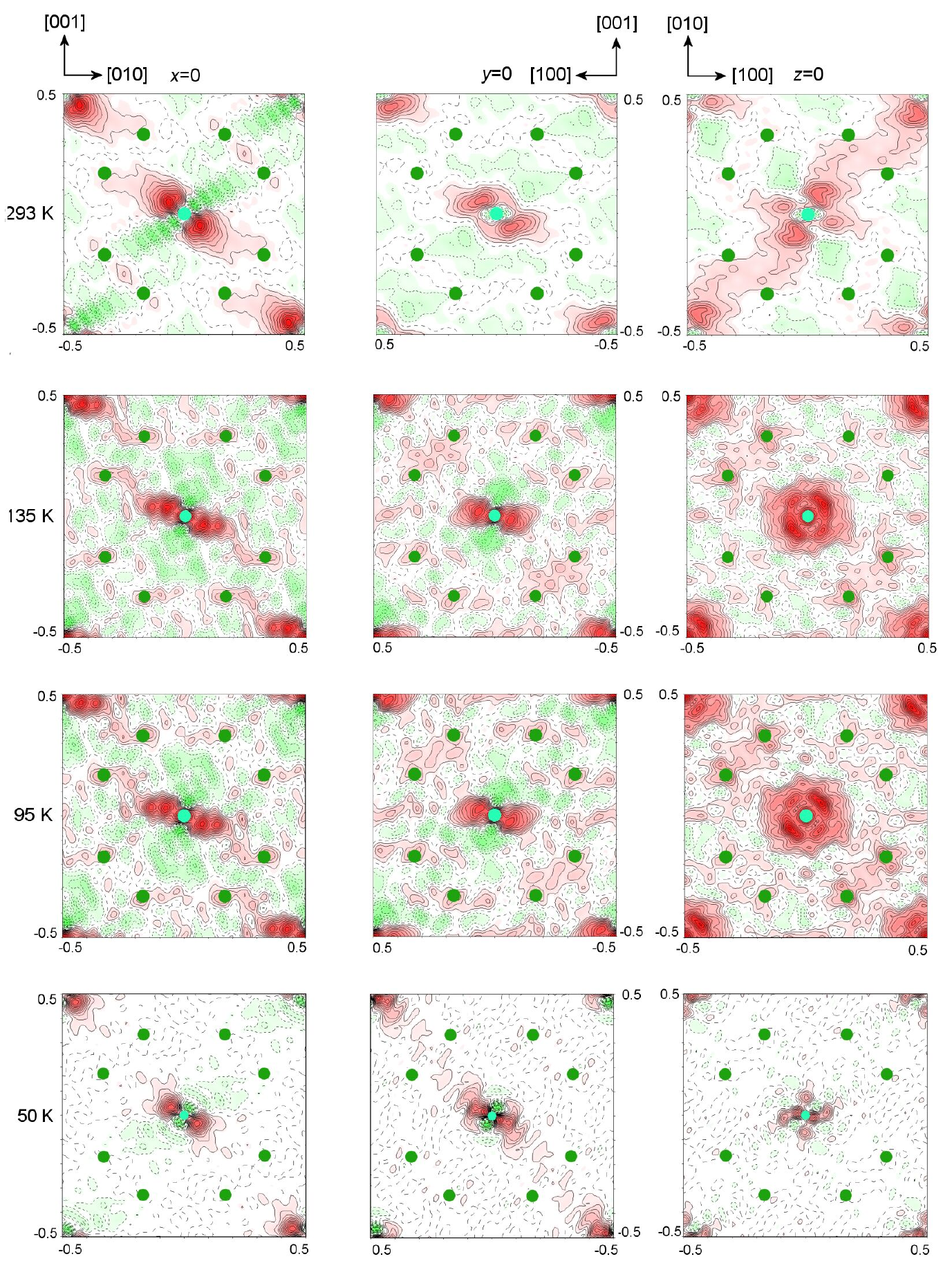}
\index{LuB$_{12}$!electron density}
\caption{Residual electron-density distribution in the $x=0$, $y=0$, and $z=0$ planes of LuB$_{12}$. Difference Fourier synthesis is done in $F\overline{1}$ using data collected at four temperatures. Contour intervals are 0.2 $e$/\AA$^3$ (295, 135, 95~K) and 1 $e$/\AA$^3$ (50~K). Positive (pink) and negative (light-green) residual electron density is highlighted. The central Lu(0,\,0,\,0) site (lime green circle) is surrounded by eight boron sites (dark green circles); $[-0.5, 0.5]$ intervals are periods of the crystal lattice \cite{Bol_BolotinaDudka18}.\vspace{-1em}}
\label{Fig:LuB12-ResidualED}
\end{figure}

\begin{figure}[t!]
\includegraphics[width=\textwidth]{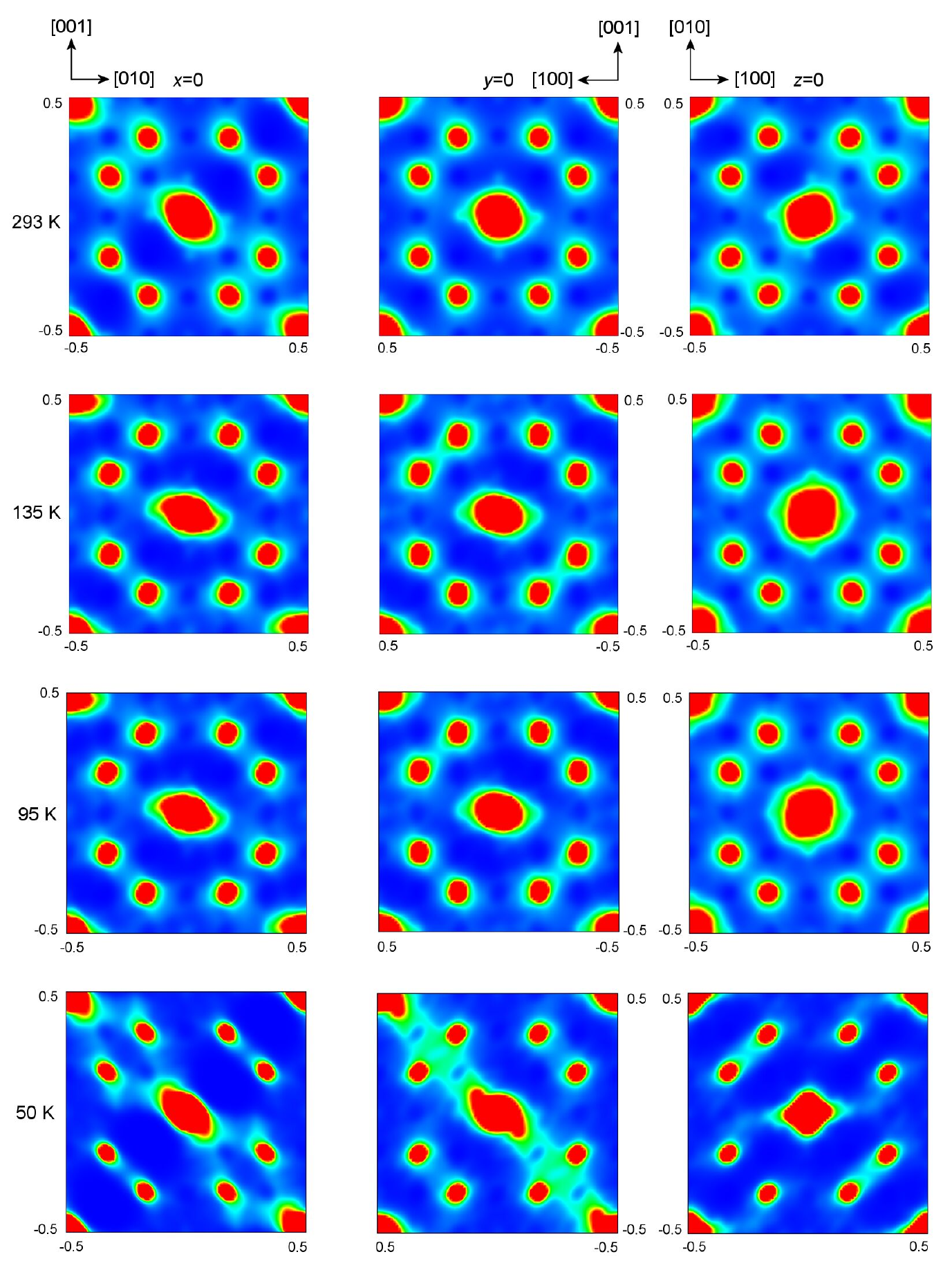}
\caption{Maximum-entropy-method maps are calculated from the LuB$_{12}$ data sets collected at temperatures 293, 135, 95 and 50~K. Three columns from left to right present thin slices of the electron-density distribution in three planes of the crystal lattice. The central Lu is surrounded by eight boron atoms; $[-0.5, 0.5]$ intervals are periods of the crystal lattice~\cite{Bol_BolotinaDudka18}.}
\label{Fig:LuB12-MEM-maps}
\end{figure}

In the case when accuracy and reliability of measured x-ray data are ensured by reliable measurement of literally each reflection, with subsequent consideration of experimental corrections using special techniques \cite{Bol_Dudka07, Bol_Dudka10, Bol_Dudka18}, one may feed a less symmetrical group to the input of the Fourier procedure. In \cite{Bol_SluchankoBogach18}, this approach was applied to LuB$_{12}$ whose structure was first refined in the high-symmetry $Fm\overline{3}m$ group at temperatures 295 and 90~K. After that, the measured values of $|F_{\rm obs}|$ were averaged in the $mmm$ Laue class instead of $m\overline{3}m$, and the orthorhombic $Fmmm$ group was fed to the input of the difference Fourier procedure, skipping the tetragonal $I4/mmm$ group, which would require a transition to another unit cell. The difference Fourier maps built from low-temperature (90~K) x-ray data clearly showed residual electron-density peaks oriented along $[001]$ at distances of about 0.5~\AA\ from the central position of Lu. As can be seen from Fig.~\ref{Fig:LuB12-DiffFourier}, similar peaks are absent along $[010]$, which is symmetrically equivalent to $[001]$ in the cubic group. This result agrees with the result obtained in the same work \cite{Bol_SluchankoBogach18} concerning unequal magnetoresistance in the LuB$_{12}$ sample in two directions of the $\langle 100 \rangle$ family.

In the next work \cite{Bol_BolotinaDudka18}, the crystal structure of LuB$_{12}$ was studied at the four temperatures 293, 135, 95 and 50~K. To eliminate possible dependence of the results on systematic instrumental errors and on the features of the crystalline sample, the x-ray experiments were performed on three different-type diffractometers and on two LuB$_{12}$ crystals. To analyze the electron-density distribution in the crystal at room temperature, the same data were used that were previously collected on a CAD4 diffractometer (Enraf Nonius) for a precise analysis of the cubic structure of LuB$_{12}$ \cite{Bol_DudkaKhrykina17}. The x-ray data at 135 and 95~K were collected on an Xcalibur EOS S2 diffractometer with a two-dimensional CCD detector. The experiment at 50~K was obtained on a four-circle Huber-5042 diffractometer equipped with a point detector and a closed-cycle helium cryostat Displex DE-202. The structure was first refined in $Fm\overline{3}m$ as before, but information on triclinic $F\overline{1}$ symmetry was fed to the input of the Fourier procedure. Non-standard abbreviation $F\overline{1}$ instead of $P\overline{1}$ is due to the reluctance to move to another (non-cubic) cell, which would correspond to the standard setting. The difference Fourier maps were built for each temperature in three sections of a crystal with the (100), (010), (001) planes. As seen from Fig.~\ref{Fig:LuB12-ResidualED}, the symmetry of the residual electron-density distribution is clearly lower than orthorhombic. The selected directions remain but lose their exact orientation along the canceled axis 2 of the orthorhombic group, turning in the direction closer to the face diagonal of the unit cell. The residual electron density increases almost by an order of magnitude at the temperature of 50~K forming a continuous diagonal strip in the (010) section. The formation of the electron-density strip at 50~K is confirmed by the maximum entropy method (MEM) as shown in Fig.~\ref{Fig:LuB12-MEM-maps}.

\begin{figure}[b!]
\begin{center}
\includegraphics[width=\textwidth]{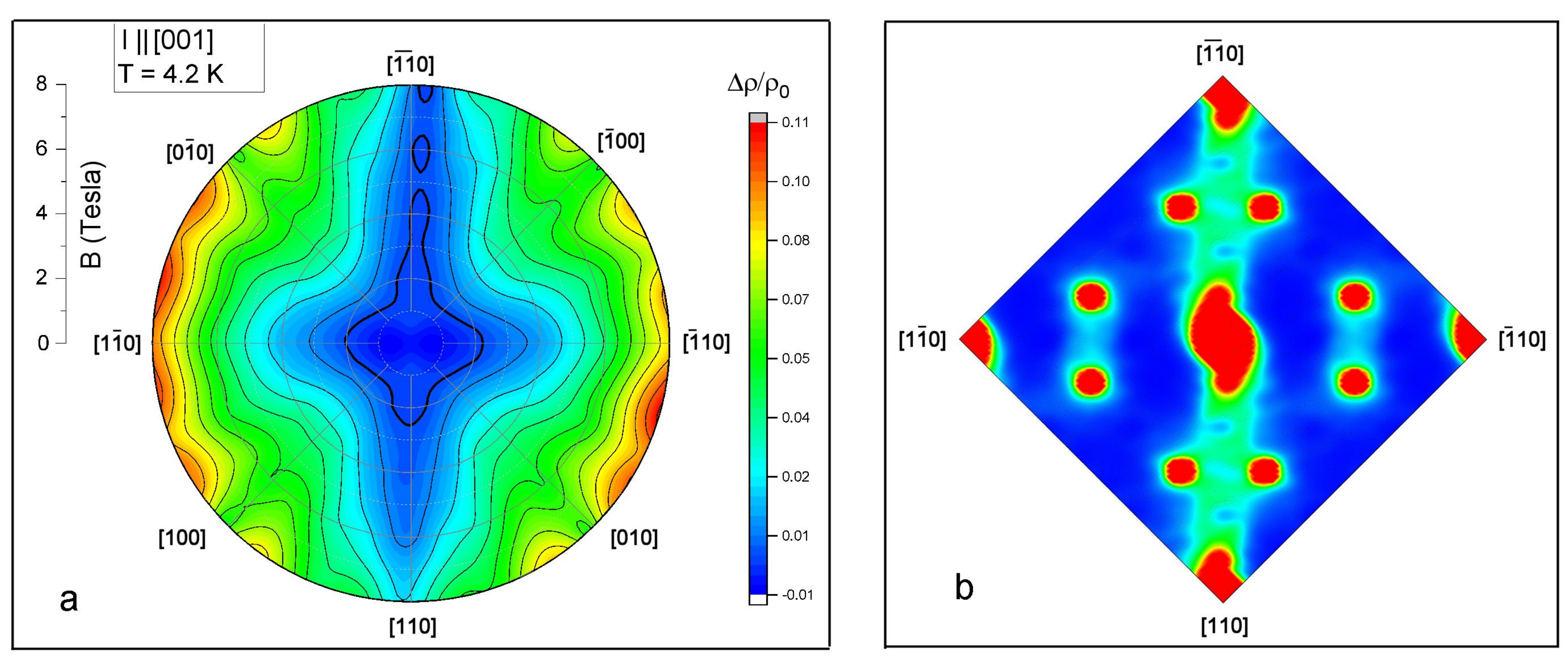}\vspace{-3pt}
\end{center}\index{LuB$_{12}$!magnetoresistance anisotropy}\index{LuB$_{12}$!electron density}
\caption{(a)~Magnetoresistance anisotropy of LuB$_{12}$ in polar coordinates: $\Delta\rho/\rho_0=[\rho(\varphi,B)-\rho(\varphi_0,B)]/\rho(\varphi_0, B)$, \mbox{$\varphi_0 = 270^\circ$} corresponding to $\mathbf{B} \parallel [\bar{1}\bar{1}0]$; (b)~anisotropic electron-density distribution in a thin layer of the electron density reconstructed by the maximum entropy method~\cite{Bol_BolotinaDudka18}.\vspace{-1pt}}
\label{Fig:LuB12-Magnetoresistance}
\end{figure}

We associate this observation with the formation of a filamentary structure of conductive channels\,---\,charge stripes along selected directions in the crystal \cite{Bol_BolotinaDudka18}. In the same paper, two results were compared, which were obtained on LuB$_{12}$ samples cut from one block. The same sample could not be used in all experiments due to different requirements for its size and shape for x-ray experiments and measurements of transport and magnetic properties. Moreover, the x-ray measurements were carried out at significantly higher temperatures and in the absence of an external magnetic field. The more surprising is the exact orientational coincidence of two pictures in the left and right parts of Fig.~\ref{Fig:LuB12-Magnetoresistance}, one of which (left) illustrates the anisotropy of the transverse magnetoresistance in LuB$_{12}$ whereas the second picture demonstrates the anisotropy of the residual electron-density distribution in LuB$_{12}$ at 50~K.

\begin{figure}[b!]
\begin{center}\vspace{-2pt}
\includegraphics[width=\textwidth]{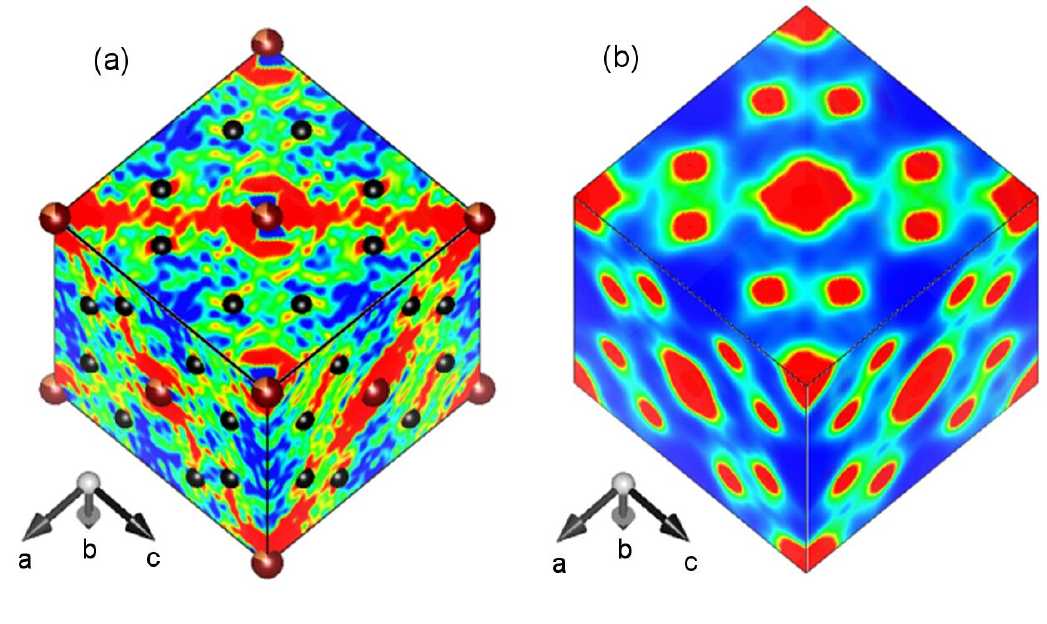}\vspace{-3pt}
\end{center}\index{Tm$_{1-x}$Yb$_x$B$_{12}$!electron density}
\caption{(a)~Difference Fourier and (b)~maximum-entropy-method maps of Tm$_{0.19}$Yb$_{0.81}$B$_{12}$ are created in (100), (010), (001) faces of the unit cell. Electron density ($g$) in the layer of any given thickness is automatically divided into several levels from $g_{\rm min}$ to $g_{\rm max}$, each of them is assigned to a definite color from dark-blue over green to red. The values of $g_{\rm MEM}$ are cut at the level $g_{\rm max} = 0.075$\% of the maximal $g_{\rm MEM}$ value to show fine electron-density gradations in the thin layer. Difference electron-density values are cut at $\pm 0.5$~$e$/\AA$^3$~\cite{Bol_SluchankoAzarevich19}.\vspace{-1pt}}
\label{Fig:TmYbB12-DifMEM}
\end{figure}

Another structure of a single-crystal Tm$_{0.19}$Yb$_{0.81}$B$_{12}$ was analyzed according to the same scheme at room temperature~\cite{Bol_SluchankoAzarevich19}. Extreme members TmB$_{12}$ and YbB$_{12}$ in a series of solid solutions Tm$_{1-x}$Yb$_x$B$_{12}$ vary greatly in their properties, despite the proximity of Tm and Yb in the series of rare-earth elements. Unlike metallic TmB$_{12}$ with antiferromagnetic properties, YbB$_{12}$ is a narrow-gap semiconductor known as a Kondo insulator. In order to analyze the loss of metallic properties when thulium is replaced by ytterbium, the information is needed on the corresponding changes in the crystal structure.

It has been determined that the crystal lattice of Tm$_{0.19}$Yb$_{0.81}$B$_{12}$ has the same type of distortion as that of LuB$_{12}$, with $a\approx b > c$ and a small difference of about 0.002~\AA\ between the smaller lattice constant and the other two. The residual electron density is oriented predominantly along the three face diagonals of the unit cell. They are connected by a spatial diagonal, which is one of the three-fold axes of the undistorted cubic structure. The residual electron density forms a strip along one of the face diagonals even at room temperature, as can be seen in Fig.~\ref{Fig:TmYbB12-DifMEM}.\vspace{-1pt}
\index{charge stripes|)}

\section{Conclusions}

The results presented in this chapter demonstrate the complexity of the atomic structure of the dodecaborides, a complete description of which does not fit into the framework of a simple cubic model. Both atomic coordinates being expressed in fractions of the lattice constants and ADPs of almost all dodecaborides correspond well to cubic symmetry and do not require revision of the structural model despite the Jahn-Teller distortion of lattice parameters. Symmetry violations manifest themselves in difference Fourier syntheses as an asymmetric distribution of the residual electron density in the interstices of the crystal lattice along symmetrically equivalent directions. These results are in good agreement with the observed asymmetry of physical properties (conductive, magnetic).

Another prospective direction of the structural analysis of dodecaborides is the quantitative analysis of the temperature behavior of the atomic displacement parameters using multi-temperature x-ray data. The dynamics of the crystal lattice can be traced without going beyond the cubic structural model, by matching the equivalent atomic displacement parameters to the extended Einstein or Debye models.

The analysis of the dodecaboride structure is thus not limited to the refinement of the structural model in the high-symmetric group at one or several temperatures. The multi-temperature data on ADPs must be supplemented with the temperature dependent unit-cell parameters, which are not bound by the symmetry constraints, and with the difference Fourier maps built without reliance on the symmetry of the structure model.

The transition from single experiments to systematic research of the structure-property relationship in dodecaborides requires the creation of a database of diffraction data. For reliable characterization of a single dodecaboride of a certain composition, it is necessary to carry out a series of diffraction experiments in a wide temperature range with the maximum possible coverage of the low-temperature region. The temperature step should be selected individually for each composition in order to monitor the structural parameters.

\section*{Acknowledgments}
\addcontentsline{toc}{section}{Acknowledgments}

The authors are grateful to N.~E. Sluchanko and N.~Yu. Shitsevalova for useful discussions. This work was supported by the Ministry of Science and Higher Education within the state assignment of the Federal Scientific Research Center (FSRC) ``Crystallography and Photonics'' of the Russian Academy of Sciences in the part related to the development of structural analysis methods. Crystal structures and properties of HoB$_{12}$ and ErB$_{12}$ crystals were studied with the support of the Russian Foundation for Basic Research, Grant No.~18-29-12005; similar studies on TmB$_{12}$, YbB$_{12}$, and LuB$_{12}$ were supported by the Russian Science Foundation, grant No.~17-12-01426. The diffraction data were collected using the equipment of the Shared Research Center of the FSRC ``Crystallography and Photonics'' of the Russian Academy of Sciences and was supported by the Russian Ministry of Education and Science (project RFMEFI62119X0035).


\setcounter{section}{0}
\putbib[Chapter_Bolotina]

\end{bibunit}

\printindex

\end{document}